\documentclass[fleqn,usenatbib]{mnras}

\usepackage{newtxtext,newtxmath}
\usepackage{booktabs}
\usepackage{svg}
\usepackage{lastpage}

\usepackage[T1]{fontenc}
\usepackage{xcolor}
\usepackage{color}

\DeclareRobustCommand{\VAN}[3]{#2}
\let\VANthebibliography\thebibliography
\def\thebibliography{\DeclareRobustCommand{\VAN}[3]{##3}\VANthebibliography}

\usepackage{graphicx}	
\usepackage{amsmath}	
\usepackage{makecell}

\title[1D vs. 3D atmospheric modelling]{Quantifying the differences in transmission and emission spectra of hot irradiated gaseous exoplanet atmospheres: A comparison of 1D and 3D modelling using JWST}

\author[R. Arora and L. Majumdar]{Rahul Arora,$^{1,2}$, Liton Majumdar,$^{1,2}$\thanks{E-mail: liton@niser.ac.in; dr.liton.majumdar@gmail.com}
\\
$^{1}$Exoplanets and Planetary Formation Group, School of Earth and Planetary Sciences, National Institute of Science Education and Research, 
Jatni 752050, Odisha, India\\
$^{2}$Homi Bhabha National Institute, Training School Complex, Anushaktinagar, Mumbai 400094, India\\}

\date{Accepted XXX. Received YYY; in original form ZZZ}

\pubyear{\the\year{}}

\begin{document}
 
\label{firstpage}
\pagerange{\pageref{firstpage}--\pageref{LastPage}}\maketitle

\begin{abstract}

Modeling the atmospheres of exoplanets is fundamental to understanding their atmospheric physics and chemical processes. While one-dimensional (1D) atmospheric models with 1D radiative transfer (RT) have been widely used, advances in three-dimensional (3D) general circulation models (GCMs) and 3D RT methods now allow quantitative comparisons of these approaches. With the precision and sensitivity of JWST, such differences can be observationally tested. This study investigates the spectral variations produced by 1D and 3D models and estimates the JWST observing time or number of transits needed to distinguish them. Using HD 189733b as a case study, three sets of simulations were performed: 1D atmospheric models with 1D RT and 3D GCM models coupled with both 1D and 3D RT. An inherent limitation of our study is that the temperature–pressure (T–P) profiles derived from the 3D GCM extend only to the high-pressure regions. The simulations incorporated both equilibrium and disequilibrium chemistry. Significant spectral discrepancies were found, with 3D models generally showing weaker features. Using a JWST noise simulator, the signal-to-noise ratio (SNR) for detecting these differences was calculated. For transmission spectra, the SNR ranged from 2.04–7.68 (equilibrium) and 1.66–7.04 (disequilibrium), while for emission spectra it ranged from 5.90–34.52 (equilibrium) and 7.11–36.93 (disequilibrium). To test the limitations of the 3D GCM, we extended the atmosphere to lower pressures using an isothermal T–P profile and found wavelength-dependent variations in both the spectra and the SNR. These results show that JWST can distinguish 1D from 3D model spectra for major molecular features, underscoring the importance of 3D modeling in interpreting exoplanetary atmospheres.

\end{abstract}




\section{Introduction} \label{sec:intro}

Since the discovery of the first exoplanet in 1992, a key focus in astrophysics has been the discovery of more exoplanets and the understanding of their atmospheric characteristics \citep{seager10}. The characterization of exoplanetary atmospheres has significantly advanced through the development of atmospheric models \citep{JF18}. These models establish crucial links between the physical and chemical processes occurring on a planet and the observed photons, allowing us to infer the properties of the atmosphere. As a result, they contribute to our understanding of key planetary processes, including temperature and pressure profiles, as well as the specific chemical composition of the atmosphere \citep{madhu2019, JF21}. Moreover, atmospheric models play a pivotal role in confirming the presence of key atmospheric gases, such as water vapor, carbon dioxide, methane, and others, by identifying their spectral signatures \citep{madhu2019}. The ability of these models to constrain the abundance of gases is essential for determining a planet's habitability, evolutionary history, and potential for hosting life \citep{madhu24}.\par

Nowadays, the most widely used atmospheric models for exoplanets are one-dimensional, modeling the physical structure of the exoplanet in the vertical direction. Some of the most commonly used one-dimensional Radiative-Convective Equilibrium (RCE) models in the community include HELIOS \citep{Helios}, ATMO \citep{ATMO1, ATMO2}, and PICASO 3.0 \citep{PICASO3.0}. These are accompanied by one-dimensional radiative transfer models such as petitRADTRANS \citep{petit}, PICASO \citep{picaso}, and PHOENIX \citep{Phoenix_rt}. Although these models have been widely used together with several JWST space-based observations (with \citet{ers23} being among the most recent, where multiple such forward models were used), they have a few limitations, such as assuming local thermodynamic equilibrium (LTE) and hydrostatic equilibrium, with the most significant limitation being the assumption of a one-dimensional atmosphere. These models rely on the assumption of horizontal homogeneity of planetary properties, such as temperature, pressure, and chemical composition, including the volume mixing ratios of all gases. \par

To overcome the assumptions of one-dimensional models, two-dimensional models (vertical and one horizontal direction), such as PLATON \citep{2019ascl.soft03014Z} and ATMO \citep{Tremblin2015, Drummond2016, Goyal2018}, as well as three-dimensional models like SPARC/MITgcm \citep{Showman_2009}, THOR \citep{THOR+HELIOS}, Exo-FMS \citep{exofms}, and the UK Met Office's model \citep{UM}, have been developed. These models are primarily derived from existing three-dimensional (3D) General Circulation Models (GCMs) developed for Earth. Specifically, 3D GCMs have been extensively used to model a diverse range of exoplanets, including hot Jupiters such as HD 80606b \citep{806063d}, HD 189733b \citep{hd189733b3d}, and WASP-39b \citep{Wasp39b3d}, as well as terrestrial planets like TRAPPIST-1c \citep{trap3d} and sub-Neptunes such as K2-18b \citep{k2183d}. Multiple studies have also analyzed the 3D distribution of atmospheric chemistry, focusing on particular aspects of disequilibrium chemistry \citep{drummond2020,2023A&A...672A.110L,zam2023}. While these models provide a more comprehensive and higher-dimensional representation of a planet's properties, they are computationally expensive, and it remains extremely challenging to study their dynamical, radiative, and chemical processes self-consistently in 3D \citep{Heng15}. This raises the question: Can 3D GCMs provide better constraints on planetary physics and chemistry, such as temperature-pressure profiles and atmospheric compositions, from observed transmission and emission spectra compared to 1D models? Additionally, do the model spectra from 3D GCMs show significant differences when compared to those from 1D models that could be observed by current telescopic facilities like the James Webb Space Telescope (JWST)?
\begin{figure*}
\centering
\includegraphics[width=0.6\linewidth]{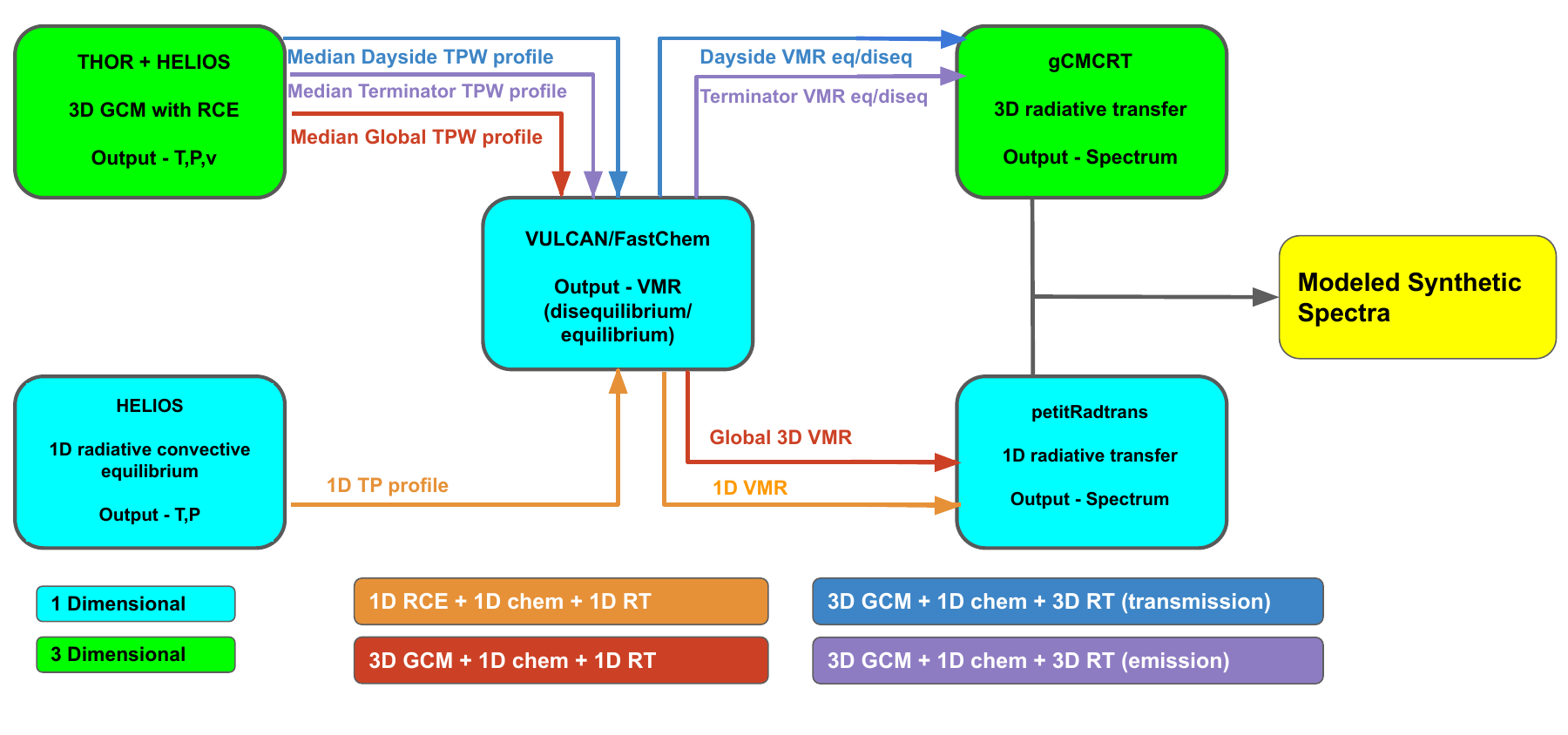}
\caption{Flowchart illustrating our methodological pipeline, which couples atmospheric, chemical, and radiative transfer codes for each type of simulation. The color codes are as follows: A1 and A2 in orange, B1 and B2 in red, transmission cases C1 and C2 in blue, and emission cases C1 and C2 in light purple. These cases are discussed in Table \ref{table:modelcodes}.}
\label{fig:flow}
\end{figure*}

To address this question, we modeled HD 189733b, the closest hot Jupiter to Earth at a distance of 63 light-years, and compared the modeled synthetic spectra derived from the 1D RCE model, HELIOS \citep{Helios}, and the 3D GCM, THOR \citep{THOR+HELIOS}, coupled with the 1D radiative transfer code petitRADTRANS \citep{petit} and the 3D radiative transfer code gCMCRT \citep{gCMCRT}. We employed both equilibrium chemistry, using FASTCHEM \citep{Fastchem}, and disequilibrium chemistry, using VULCAN \citep{VULCAN}. Our source selection is based on the fact that HD 189733b is one of the most well-studied hot Jupiters, particularly suited for atmospheric characterization due to its large size, low mean molecular weight, and large scale heights. Substantial temperature differences between the day and night sides make it an ideal candidate for studying the effects of 3D GCM modeling as well \citep{Heng15}. Observations of its transmission and emission spectra using the Hubble Space Telescope (HST) and ground-based facilities have confirmed the presence of several atoms and molecules, including H$_2$O \citep{HD189water1,HD189water2}, CO \citep{HD189co}, Na \citep{HD189na}, HCN \citep{HD189hcyn}, and K \citep{HD189k}. Recently, JWST confirmed the detections of H$_2$O and CO and identified new molecules, such as CO$_2$ and H$_2$S \citep{fu2024hydrogensulfidemetalenrichedatmosphere,2024ApJ...973L..41I}. Additionally, a 3D thermal profile of HD 189733b using Spitzer/IRAC at 8 $\mu$m \citep{3dhdspitzer} is also available.

The paper is organized as follows: In Section \ref{sec:methods}, we describe the procedures used to generate the temperature-pressure profiles of the 1D RCE and 3D GCM models, their coupling with chemical codes, and their integration with radiative transfer codes to model transmission and emission spectra under both equilibrium and disequilibrium conditions. Additionally, we describe the noise calculation using the state-of-the-art noise simulator for JWST, PANDEXO \citep{Pandexo}. In Section \ref{sec:results}, we present the spatial variation of physical quantities, show the temperature-pressure profiles and their differences between the 1D and 3D models, along with the differences in volume mixing ratios of major species in the atmospheres under both equilibrium and disequilibrium conditions, and finally present the model transmission and emission spectra. We also discuss the results from our noise calculation and their potential implications for future JWST observations in understanding the 3D nature of the atmospheres as compared to the 1D case. Finally, in Section \ref{sec:conc}, we summarize our findings and conclusions.

\section{Methodology} \label{sec:methods}
 Our methodological pipeline for assessing the quantified differences between various combinations of 1D and 3D modeling frameworks and observable spectra is outlined in Figure \ref{fig:flow}. In this section, we present a detailed description of the methods followed.

\subsection{Physical Structure of HD189733b}

HD189733b is the nearest Hot Jupiter to Earth and one of the most extensively studied exoplanets. It has a radius of 1.126 \(R_J\) and orbits a K-type star with a radius of 0.805 \(R_\odot\) and a temperature of 4875 K \citep{Boyajian_2014}. To generate the physical structure of HD189733b, we computed the variation of temperature and the \(K_{ZZ}\) coefficient/wind velocities with respect to pressure/height in the atmosphere using both 1D and 3D models.

For the 1D model, we used the self-consistent radiative transfer code HELIOS \citep{Helios}, which solves for radiative-convective equilibrium to constrain the temperature-pressure profile in the vertical direction. In contrast, the temperature-pressure (T-P) variation in 3D was modeled using THOR + HELIOS, a General Circulation Model (GCM) (Version 2) with a multiwavelength radiative transfer code \citep{THOR+HELIOS}. For radiative transfer calculations, both the 1D and 3D GCM models considered H$_2$O, HCN, CH$_4$, CO, CO$_2$, NH$_3$, and H$_2$-H$_2$, as well as H$_2$-He Collision-Induced Absorption (CIA) as sources of opacities . For our simulation, we used a grid of 104 vertical layers spanning a pressure range from 10$^{-8}$ to 200 bar for the 1D model with HELIOS, and we extended the THOR output to the same pressure range using an isothermal T-P profile to ensure consistency between the 1D and 3D GCM grids obtained from THOR.

\begin{table}
\caption{Input parameters used to simulate the 3D atmosphere of HD189733b using the THOR GCM.}
\label{table:thor_parameters}
\begin{tabular}{@{}ll@{}}
\toprule
\toprule
\textbf{Parameters}               & \textbf{Value}             \\ \midrule
Timestep                 & 30 s              \\
Total steps              & 2,880,000         \\
Radius                   & 786,498,750 m    \\
Rotation rate            & $3.28 \times 10^{-5}$ rad/s \\
Gravitation              & 21.5 m/s$^2$      \\
Gas constant             & 3516.1 J/(kg K)   \\
Specific heat capacities & 12,500 J/(kg K)   \\
Mean temperature         & 1162 K            \\
Reference pressure       & 200 bar           \\
Top altitude            & 3,000,000 m       \\
GCM grid subdivision     & 4                  \\
Vertical layers          & 53                 \\
Stellar temperature      & 4875 K            \\
Star-planet distance    & 0.03142 au        \\
Stellar radius           & 0.805 R$_{\odot}$ \\
Radiative transfer step  & 2                  \\ \bottomrule
\end{tabular}
\end{table}

Table \ref{table:thor_parameters} lists the parameters used in the GCM model, which was run under the assumption of non-hydrostatic equilibrium with the reference base pressure set at 200 bar \citep{Deitrick_2020}. The simulation grid was chosen to be an icosahedron with 2562 vertices. The model evolved the planet for 1000 Earth days by solving Euler's equations with a timestep of 30 s and performing radiative transfer calculations every five timesteps (i.e., every 150 s). After a few hundred days, the changes in temperature, pressure, and winds became negligible, and the model converged.

In the case of the 3D model, the wind profile from the GCM simulation is used to calculate \(K_{ZZ}\), following \cite{Deitrick_2020}. However, in the case of the 1D model, due to a lack of wind information, a parametric profile was used, given by Equation \ref{eq:KZZ_1D}, following \cite{VULCAN}:
\begin{equation}
    K_{ZZ} = K_{deep} \left(\frac{P_{trans}}{P}\right)^{0.4}
    \label{eq:KZZ_1D}
\end{equation}
where \(K_{deep}\) is the eddy diffusion coefficient in the deeper regions of the planet, \(P_{trans}\) is the transition pressure level, and \(P\) is the pressure of the layer.

\subsection{Chemical Variation and Observable Spectrum}

We studied the chemical variation of HD189733b under equilibrium conditions using FASTCHEM, a gas-phase equilibrium chemistry solver \citep{Fastchem}, and under disequilibrium conditions using VULCAN, a Python-based chemical kinetics code \citep{VULCAN} (Table \ref{table:Vul_parameters}). We used solar abundances \citep{Solar_2009} as the initial elemental abundances for both cases. For disequilibrium chemistry, we employed the NCHO network, which comprises 877 reactions and spans 69 molecules. In disequilibrium chemistry, we included molecular diffusion, eddy diffusion (\(K_{ZZ}\)), and photochemical processes. Throughout these simulations, we considered solar metallicity and a C/O ratio analogous to \citet{2023A&A...672A.110L}, where they introduced a reduced chemical network code, Mini-chem, and simultaneously modeled HD189733b in 3D. \\ 
\begin{table}
\caption{Parameters used for atmospheric and chemical simulations using 1D and 3D models for disequilibrium conditions. Column 2 represents the 1D radiative-convective equilibrium, and Column 3 represents the 3D general circulation model.}  
\label{table:Vul_parameters}
\centering
\begin{tabular}{@{}lll@{}}
\toprule
\toprule
Parameter               & 1D RCE        & 3D GCM                        \\ \midrule
R$_{star}$ (R$_{sun}$)  & 0.751         & 0.751                         \\
R$_{P}$ (cm)            & 7864987500    & 7864987500                    \\
Gravity (cm/s$^2$)      & 2150          & 2150                          \\
Orbit Radius (Au)       & 0.03142       & 0.03142                       \\
nz (Number of layers)   & 101           & 53                            \\
P$_b$ (Bottom pressure) & 200 bar       & GCM T-P profile \\
P$_t$ (Top pressure)    & 10$^{-6}$ bar & GCM T-P profile    \\
K$_{deep}$              & 10${8}$ m$^2$/s      & GCM wind profile              \\
P$_{trans}$             & 5 bar         & GCM wind profile              \\
Initial condition       & Equilibrium   & Equilibrium                   \\ \bottomrule
\end{tabular}
\end{table}

For modeling the observable transmission and emission spectra, we used a Python-based 1D radiative transfer code, petitRADTRANS \citep{petit}, and a 3D Monte Carlo radiative transfer code, gCMCRT \citep{gCMCRT}. For both models, we used H$_2$O, HCN, CH$_4$, CO, CO$_2$, NH$_3$, C$_2$H$_2$, and NO species as sources of opacities. Additionally, HD189733b, being a gas giant, is expected to have an H$_2$-He-based bulk atmosphere, which necessitates the inclusion of collision-induced absorption (CIA) due to H$_2$-H$_2$ and H$_2$-He. We used correlated-k opacities supplied with the petitRADTRANS package, calculated at R=1000, and the same opacities were used in the 3D radiative transfer code gCMCRT. \\

\begin{table*}
\caption{Table showing the parameters used to calculate opacity (Columns 1 and 2) and the simulation parameters for 3D radiative transfer (Columns 3 and 4) used in gCMCRT.}
\label{table:gcmcrt_parameters}
\centering
\begin{tabular}{@{}ll|ll@{}}
\toprule
\toprule
\multicolumn{2}{l}{Opacity parameters} & \multicolumn{2}{l}{Radiative transfer parameters}          \\ \midrule
\multicolumn{2}{l}{Correlated-K}      & Transmission limb sampling              & True                              \\
Pre-mixed opacity                & False       & g-ordinance bias           & True                              \\
Wavelength interpolation          & False       & Number of photons        & 102400                            \\
                                &            & Wavelength points        & 5115                              \\
Number of g ordinances           & 16          & Number of theta points   & 46                                \\
\multicolumn{2}{l}{CIA}             & Number of phi points     & 91                                \\
                                &            & Viewing theta (degrees) & 90                                \\
                                &            & Viewing phi (degrees)   & 180, 0 \\ \bottomrule
\end{tabular}%
\end{table*}

For gCMCRT, we used a 3D grid of 45 latitude points, 90 longitude points, and 53 vertical layers, simulated by THOR. We applied transit chord ray tracing and  g-ordinance biasing for the transmission spectrum. gCMCRT uses the convention of the substellar point being at 90$^{\circ}$ latitude (given by the parameter \(\phi\)) and 0$^{\circ}$ longitude (given by the parameter \(\theta\)). The observer for transmission was set at 90$^{\circ}$ \(\phi\) and 180$^{\circ}$ \(\theta\), which corresponds to the antistellar point. For the emission spectrum, the observer was set at 90$^{\circ}$ \(\phi\) and 0$^{\circ}$ \(\theta\), which is the substellar point. Other parameters are shown in Table \ref{table:gcmcrt_parameters}.

\subsection{Modeling HD 189733b in 3D and 1D}
\label{sec:trans_model}
\begin{table*}
\caption{Nomenclature of different model simulations}
\label{table:modelcodes}
\centering
\begin{tabular}{@{}llll@{}}
\toprule
\toprule
Model Code               & Atmospheric Structure        & Radiative Transfer  & Chemical Conditions                     \\ \midrule
A1  &    1D      & 1D   & Equilibrium                      \\
A2  &    1D      & 1D   & Disequilibrium                      \\
B1  &    3D      & 1D   & Equilibrium                      \\
B2  &    3D      & 1D   & Disequilibrium                      \\
C1  &    3D      & 3D   & Equilibrium                      \\
C2  &    3D      & 3D   & Disequilibrium                      \\ \bottomrule
\end{tabular}
\end{table*}

To study an exoplanet in 3D, we lack a 3D disequilibrium chemistry solver. For 3D modeling, we consider two cases: one with a 3D GCM and a 1D chemical profile with 1D radiative transfer, and a second with a 3D GCM, a 1D chemical profile, and 3D radiative transfer. In the first case, we take the global median of the T-P profile calculated at different latitudes and longitudes. This 1D global median profile is then used to calculate the vertical volume mixing ratios (VMRs) of molecules in both equilibrium and disequilibrium. Both the 1D VMR profile and the global median profile are used to perform 1D radiative transfer. 

For the second case, we divide the process into two sections: the transmission spectrum and the emission spectrum. For the transmission spectrum, we use the median profile of the east and west terminators to calculate the VMRs of molecules, generating spectra for both east and west terminator profiles by performing 3D radiative transfer. The final transmission spectrum is the average of the spectra from both terminators. For the emission spectrum, we take the median T-P profile of the dayside of the planet and repeat the method used for the transmission spectrum to obtain the emission spectrum. Both processes are performed for equilibrium and disequilibrium chemistry. 

In the above method, we use the median instead of the mean to convert the 3D Temperature-Pressure-Wind (T-P-W) distribution to 1D. This is because the minimum pressure on the planet’s dayside and nightside differ by two orders of magnitude. The mean average pressure cannot represent the correct distribution of pressure levels. To address this, we use the median of all quantities.

For the case of a 1D study of the atmosphere, we model the 1D T-P profile, calculate the K$_{ZZ}$ using Equation \ref{eq:KZZ_1D}, and determine the VMR in both equilibrium and disequilibrium conditions. Using the calculated VMR, we perform 1D radiative transfer to generate the model spectrum. A nomenclature of these models is shown in Table \ref{table:modelcodes}.

\subsection{Noise Simulation and SNR calculation}

Once the synthetic spectra are generated, we simulate JWST observations using PANDEXO \citep{Pandexo} to quantify the differences in observability between models A1, A2, B1, B2, C1, and C2. Using PANDEXO, we simulate NIRSpec (Near Infrared Spectrograph) and MIRI (Mid Infrared Spectrograph) observations covering the wavelength ranges of 0.7–5 $\mu$m and 5–14 $\mu$m, respectively. For NIRSpec, we simulate observations with three grisms: G140M (0.7–1.89 $\mu$m), G235M (1.7–3 $\mu$m), and G395M (2.9–5 $\mu$m), covering the full range of NIRSpec wavelengths. For MIRI, we include the MIRI MRS instrument (5–14 $\mu$m). We run noise simulations for each transit, which lasts 1.8 hours, and add 2 hours of baseline observation, totaling 3.8 hours. The stellar parameters used for the noise simulation are shown in Table \ref{table:Pandexo_sp}. PANDEXO uses these parameters to generate stellar fluxes using PHOENIX models. All these instruments share the same spectral resolution of 1000 in our simulation, consistent with our modeled spectra. An additional floor noise of 14 ppm is added for all simulations.

We adopt the methodology described in \citet{SNR} for the SNR calculation. The absolute difference between the 3D and 1D spectra, at which the noise is calculated, is determined first. Then, additional noise is computed using Equation \ref{eq:outnoise} to account for the out-of-transit baseline noise, as explained in \citet{SNR2}.

\begin{equation}
    N_{baseline} = \sqrt{\left(\frac{1}{t_{out}}\right)^{2} + \left(\frac{1}{t_{in}}\right)^{2}}
    \label{eq:outnoise}
\end{equation}

The signal-to-noise ratio at each wavelength point, $\left(\frac{S}{N}\right)_{i}$, is calculated using Equation \ref{eq:S/N}:

\begin{equation}
    \left(\frac{S}{N}\right)_{i} = \frac{\Delta f}{N_{baseline} \cdot N_{Pandexo}}
    \label{eq:S/N}
\end{equation}

where $\Delta f$ is the difference in transit depth or flux ratio between the 3D and 1D models, $N_{baseline}$ is the baseline noise calculated using Equation \ref{eq:outnoise}, and $N_{Pandexo}$ is the simulated noise from PANDEXO.

Using the JWST noise simulated with PANDEXO, we calculate the signal-to-noise ratio (SNR) achievable with the observation of one transit of HD189763b using Equation \ref{eq:S/N}. We then use Equation \ref{eq:S/Nt} to calculate the SNR achieved in $N_T$ transits.

\begin{equation}
    \left(\frac{S}{N}\right)_{T} = \sqrt{N_{T} \cdot \left(\frac{S}{N}\right)_{i}^2}
    \label{eq:S/Nt}
\end{equation}

\begin{table}
\caption{Stellar parameters used for PANDEXO simulation}
\label{table:Pandexo_sp}
\begin{tabular}{@{}ll@{}}
\toprule
\textbf{Parameter}            & \textbf{Value} \\ \midrule
Magnitude (J band)$^b$           & 6.07           \\
Temperature$^a$                   & 5052 K         \\
Reference wavelength (J band)    & 1.25 $\mu$m    \\
Metallicity$^a$                   & -0.02          \\
Surface gravity (log)$^a$         & 4.49           \\ \bottomrule
\end{tabular}\\
\footnotesize{The above parameters have been sourced from the following sources: a: \href{https://exo.mast.stsci.edu/exomast_planet.html?planet=HD189733b}{Exomast}, b: \href{https://simbad.u-strasbg.fr/simbad/sim-basic?Ident=hd+189733&submit=SIMBAD+search}{SIMBAD}}
\end{table}

\section{Results and Discussions} \label{sec:results}
 
In this section, we examine the quantitative differences between the modeled transmission and emission spectra of HD189733 b using 1D and 3D modeling techniques. We particularly highlight the importance of 3D General Circulation Models (GCMs) for accurately representing the physical characteristics of the planet and the effects of 3D radiative transfer, in contrast to 1D radiative transfer. We discuss our results under both equilibrium and disequilibrium conditions.

\subsection{Spatial variation of physical quantities in HD189733b}
\begin{figure}
    \centering
    \includegraphics[scale=0.55]{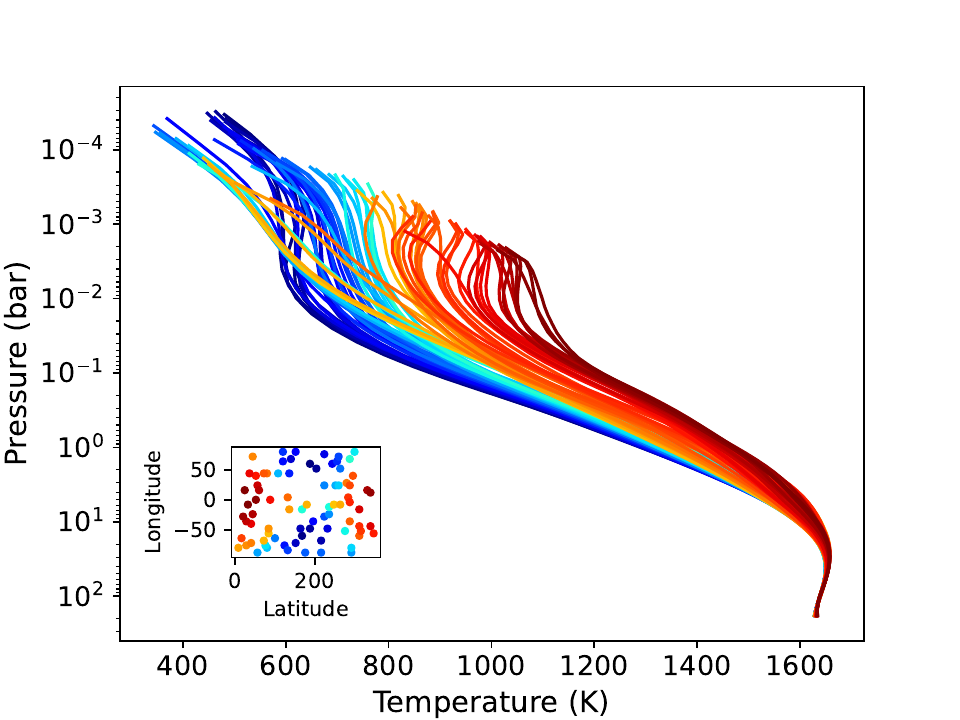}
    \caption{Comparison of T-P profiles at different latitudes and longitudes of the planet. The subplot shows the color coding of the T-P profiles with respect to latitude and longitude. The point (0,0) corresponds to the substellar point.}
    \label{fig:3DTP}
\end{figure}

We evolved the HD189733b planet for 1000 Earth days using THOR GCM, starting with a uniform distribution of the Guillot T-P profile \citep{guillot2010}. After the planet's physical processes converged, a distribution of T-P profiles was obtained, as shown in Figure \ref{fig:3DTP}. We observed an increasing variation in pressure and temperature in different parts of the planet with height.

\begin{figure}
    \centering
    \includegraphics[scale=0.55]{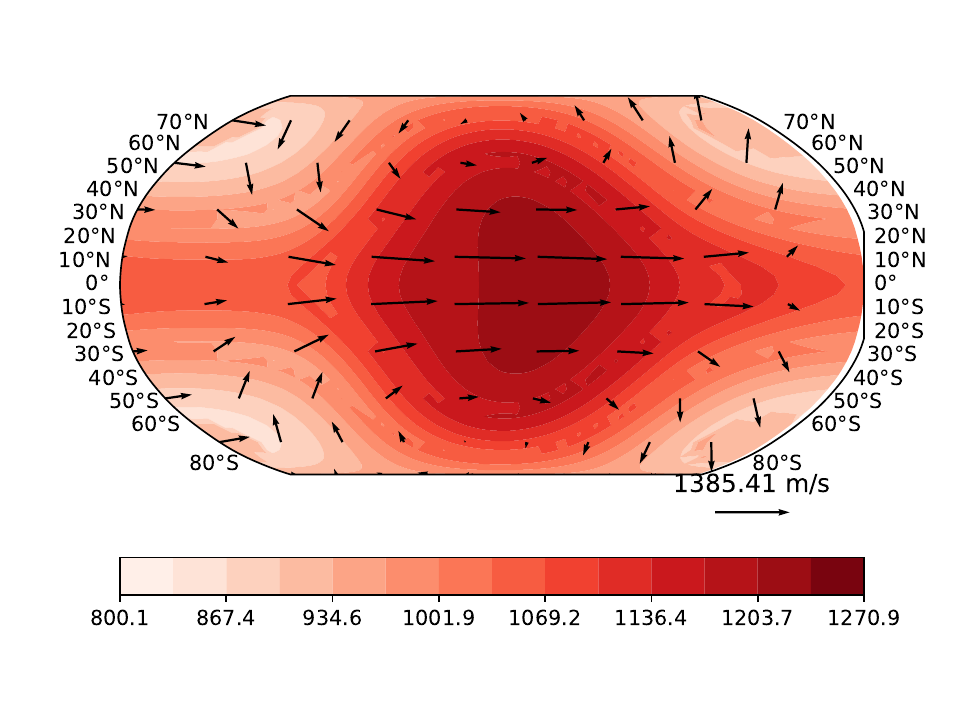}
    \caption{Horizontal distribution of temperature on the planet at a pressure level of 0.1 bar. Arrows represent the wind speed and direction at every 10° latitude and longitude point, relative to the THOR grid system.}
    \label{fig:1mbar}
\end{figure}

\begin{figure}
    \centering
    \includegraphics[scale=0.55]{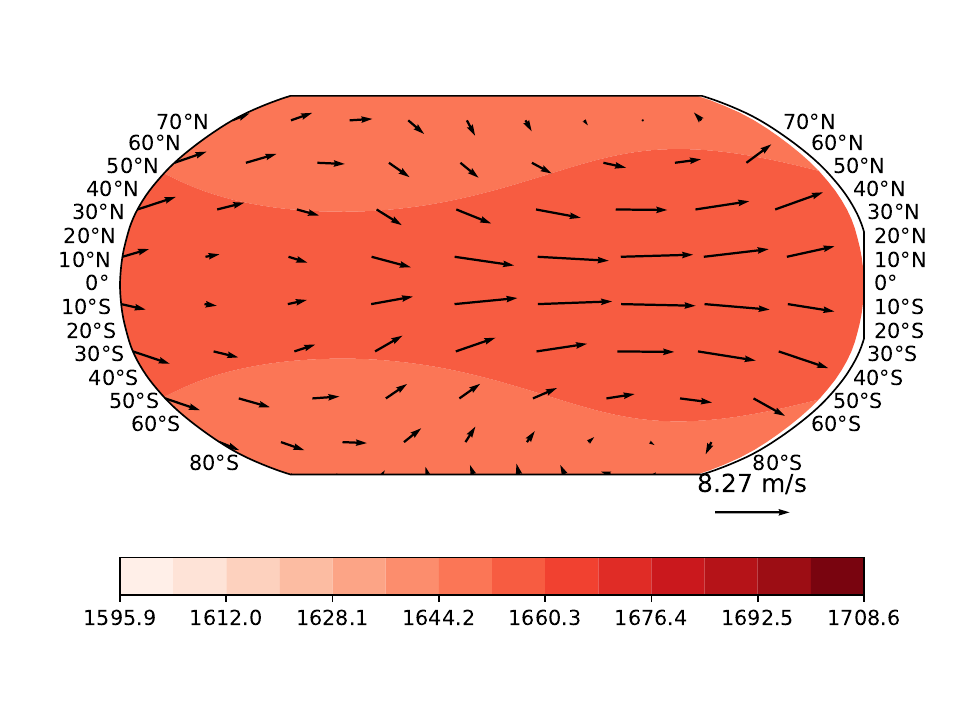}
    \caption{Same as Figure \ref{fig:1mbar}, but at a pressure level of 30 bar.}
    \label{fig:30bar}
\end{figure}

We observed that wind velocities directly impact the movement of material and, consequently, the redistribution of heat. In the lower region of the atmosphere, wind velocities are very small due to the high mass density, which makes the redistribution of material and heat inefficient, leading to a homogeneous temperature distribution, as shown in Figure \ref{fig:1mbar}. In contrast, in the upper regions of the atmosphere, high wind velocities enable efficient redistribution of heat, as can be seen in Figure \ref{fig:30bar}.

The temperature distribution of the planet showed an eastward shift of $16^{\circ} \pm 4^{\circ}$ in the planet's hotspot, due to the wind movement in the eastward direction caused by the planet's rotation from west to east. This eastward shift is also observed in Spitzer and HST data, as reported by \cite{HDhotspot2} and \cite{HDhotspot1}.

\subsection{Comparison of T-P profiles computed using 1D and 3D models}
\begin{figure}
    \centering
    \includegraphics[scale =0.55]{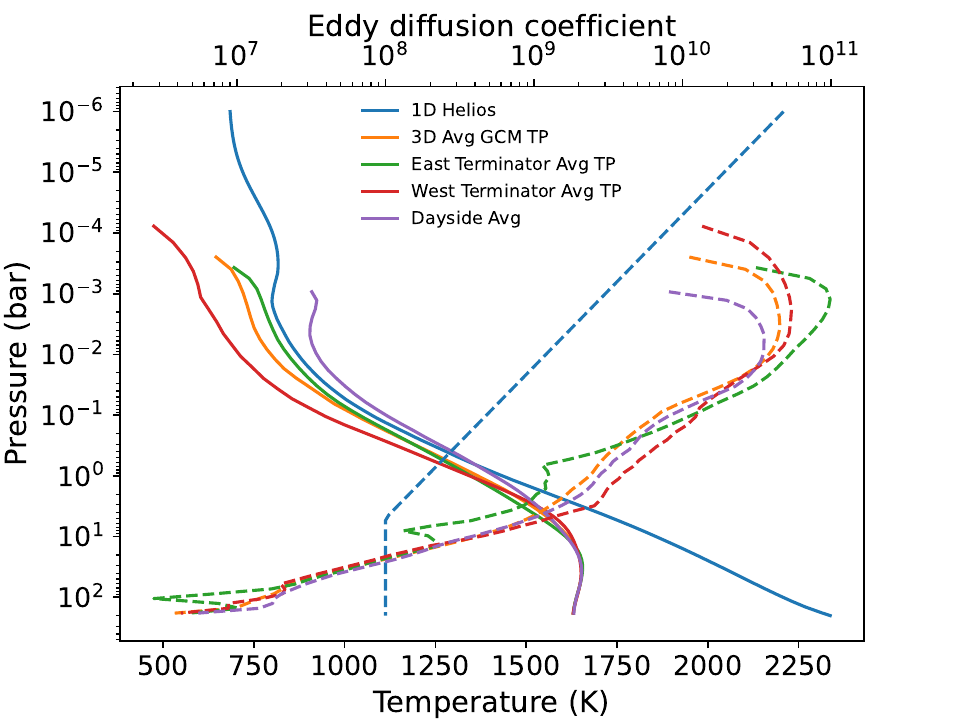}
    \caption{This figure compares the 1D radiative-convective equilibrium profile (solid blue) with the median globally averaged 3D temperature-pressure (T-P) profile (solid orange), the median east terminator average (solid green), the median west terminator average (solid red), and the median dayside average (solid purple). The dashed lines represent the eddy diffusion coefficient ($K_{ZZ}$) for each respective case.
}
    \label{fig:TP_comp}
\end{figure}

In our comparison, we generated a 1D T-P profile in radiative-convective equilibrium using HELIOS and a 3D T-P profile using the THOR GCM. For the 1D profile, we used a parametric profile for the eddy diffusion coefficient, as given by Equation \ref{eq:KZZ_1D}, with K$_{deep}$ set to $10^{8}$ cm$^{2}$s$^{-1}$ and P$_{tran}$ set to 5 bar. For the 3D GCM, the eddy diffusion coefficient profile was calculated using the wind profile obtained from the GCM output. Figure \ref{fig:TP_comp} shows the comparison between the 1D profile, the median global average 3D profile, the median east terminator average, the median west terminator average, and median dayside average  T-P and K$_{ZZ}$-P profiles.
\par

For the 1D T-P profile, we observe that the temperature is slightly higher than that of the median average T-P profile obtained from the 3D GCM. This overall shift is due to the fact that the 1D radiative-convective equilibrium assumes irradiation perpendicular to the vertical grid, while in the case of the 3D GCM, direct irradiation only affects the planet's dayside. This results in the night side being cooler, leading to a lower median average global temperature for the planet. Additionally, in the lower regions of the planet (higher pressure regions), the temperature exhibits a monotonically increasing trend in the 1D profile, whereas an isothermal nature is observed for the 3D GCM. As shown in Figures \ref{fig:1mbar} and \ref{fig:30bar}, high-pressure regions have very low horizontal wind velocity, making heat redistribution inefficient. In contrast, effective heat redistribution occurs in the low-pressure regions, preventing much stellar heat from being transferred into the lower regions of the planet, resulting in an isothermal profile. Due to wind movement from west to east, the eastern terminator ($90^{\circ}$ longitude, latitude ranging from $0^{\circ} - 90^{\circ}$) is hotter than the western terminator ($270^{\circ}$ longitude, latitude ranging from $0^{\circ} - 90^{\circ}$).
\subsection{T-P profile extension to photochemical layers}
\begin{figure}
    \centering
    \includegraphics[scale =0.55]{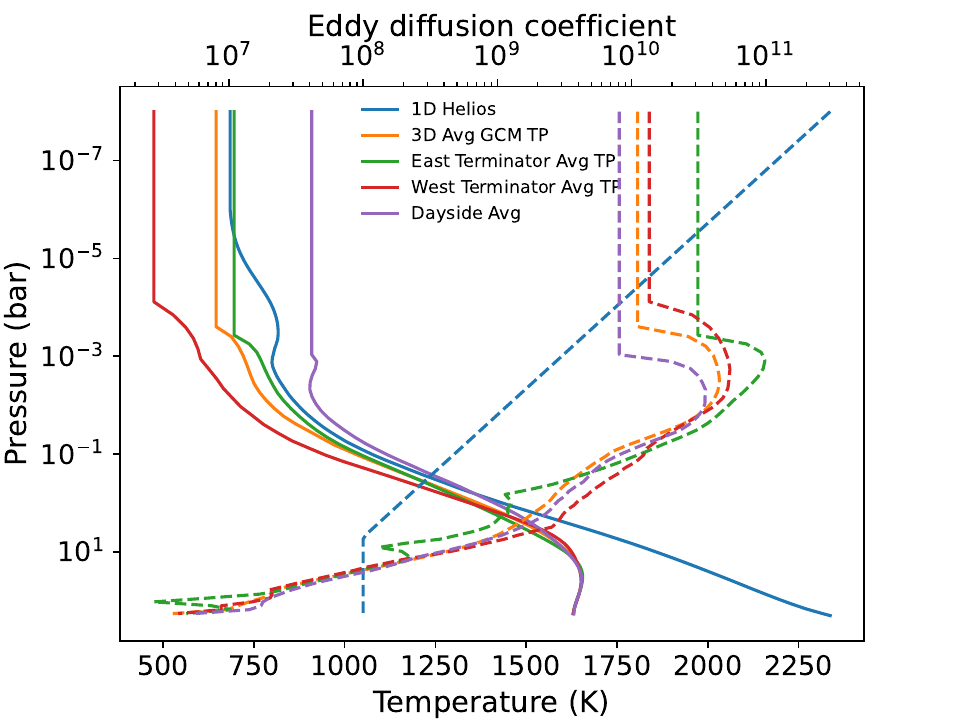}
    \caption{T–P profiles, similar to those in Figure \ref{fig:TP_comp}, are shown for an extended atmosphere ranging from 10$^{-4}$ to 10$^{-8}$ bar under isothermal conditions for all 3D cases.}
    \label{fig:TP_comp_extended}
\end{figure}

THOR is an altitude-based GCM model that does not converge at pressures lower than 10$^{-4}$~bar \citep{THOR+HELIOS}. This limitation excludes an important atmospheric region where photochemical processes are highly active. To incorporate this region, we recalculated the 1D T–P and K$_{ZZ}$ profiles with a top-of-atmosphere pressure of 10$^{-8}$~bar and extended the temperature and K$_{ZZ}$ profiles of all 3D models down to 10$^{-8}$~bar using isothermal extrapolation, ensuring consistency with the 1D profiles. Figure~\ref{fig:TP_comp_extended} compares the 1D radiative–convective equilibrium profile with the median global average 3D profile, the median east terminator average, the median west terminator average, and the median dayside average T–P and K$_{ZZ}$–P profiles, all extended from 10$^{-4}$ to 10$^{-8}$~bar under isothermal conditions.

\subsection{Variation of atmospheric volume mixing ratios across different physical structure cases}\label{subsec:VMR}
\begin{figure*}
    \centering
    \includegraphics[scale =0.45]{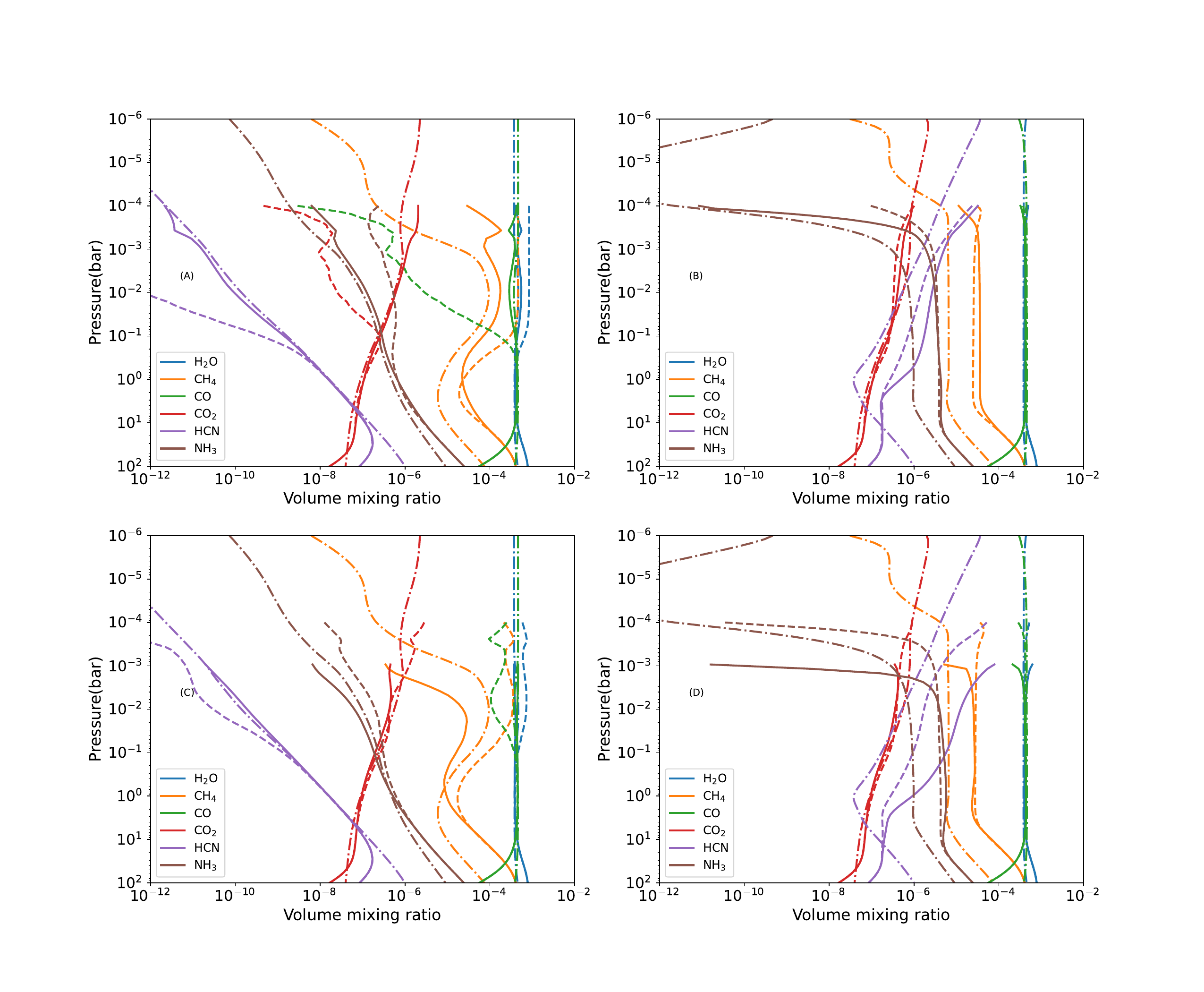}
   \caption{Vertical profiles of the volume mixing ratios for six major opacity sources under chemical equilibrium and disequilibrium conditions, calculated using the T–P and K$_{ZZ}$–P profiles shown in Figure \ref{fig:TP_comp}. Each subfigure presents results for three different temperature-pressure profiles. The first column (Panels A and C) shows equilibrium chemistry models, while the second column (Panels B and D) presents disequilibrium models. In Panels (A) and (B), the three models correspond to the median east-side terminator average from the 3D GCM (solid line), the median west-side terminator average from the 3D GCM (hashed line), and the 1D radiative-convective equilibrium model (hashed + dotted line). In Panels (C) and (D), the three models represent the median dayside average from the 3D GCM (solid line), the median global terminator average from the 3D GCM (hashed line), and the 1D radiative-convective equilibrium model (hashed + dotted line).}  
    \label{fig:vmr_eq_term}
\end{figure*}
\begin{figure*}
    \centering
    \includegraphics[scale =0.45]{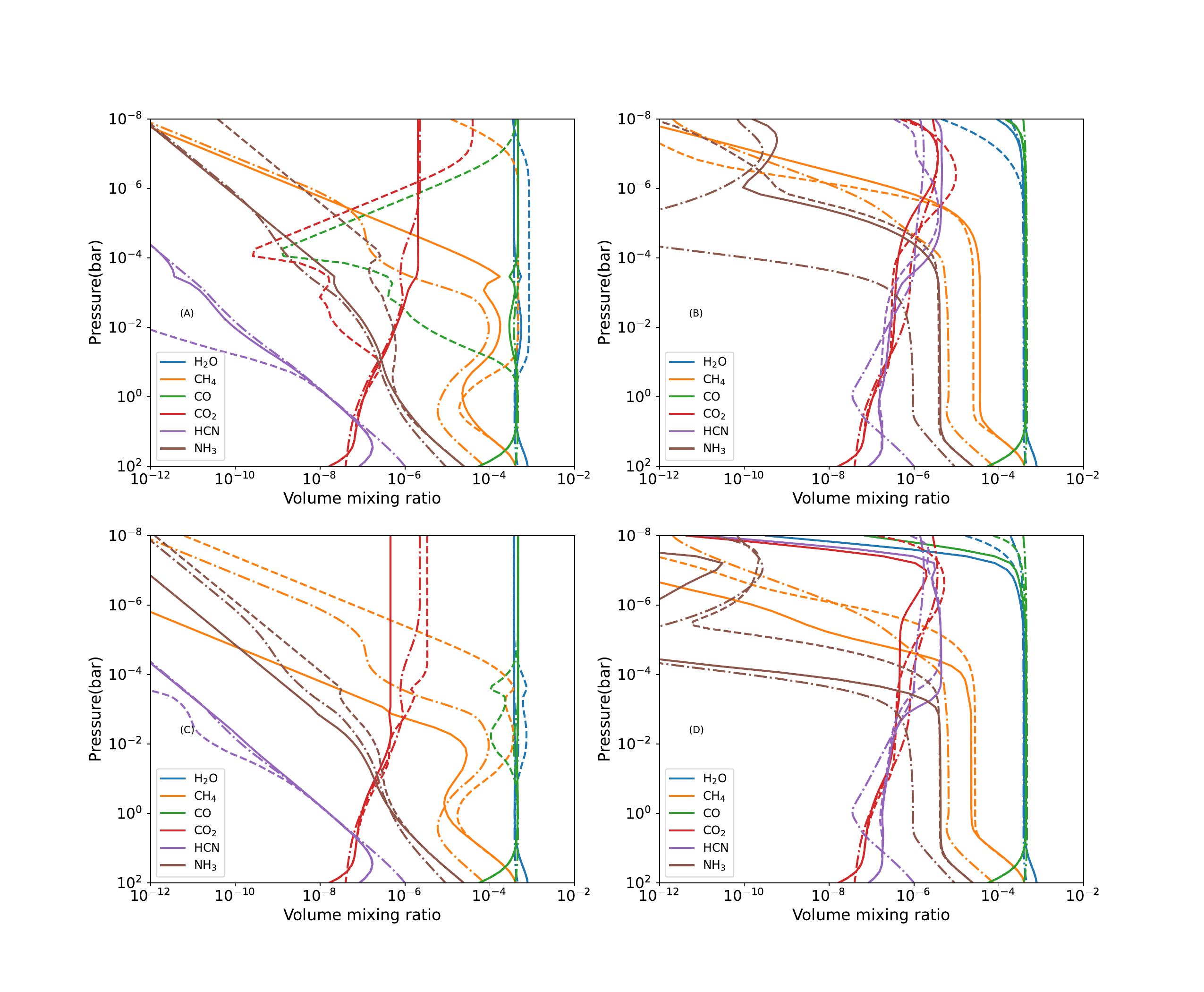}
   \caption{Vertical profiles of the volume mixing ratios for six major opacity sources under chemical equilibrium and disequilibrium conditions, calculated using the T–P and K$_{ZZ}$–P profiles shown in Figure \ref{fig:TP_comp_extended}. Each subfigure presents results for three different temperature-pressure profiles. The first column (Panels A and C) shows equilibrium chemistry models, while the second column (Panels B and D) presents disequilibrium models. In Panels (A) and (B), the three models correspond to the median east-side terminator average from the 3D GCM (solid line), the median west-side terminator average from the 3D GCM (hashed line), and the 1D radiative-convective equilibrium model (hashed + dotted line). In Panels (C) and (D), the three models represent the median dayside average from the 3D GCM (solid line), the median global terminator average from the 3D GCM (hashed line), and the 1D radiative-convective equilibrium model (hashed + dotted line).}  
    \label{fig:vmr_eq_term_extended}
\end{figure*}
We calculated the volume mixing ratios (VMRs) for five temperature–pressure–wind (T–P–W) profiles under both equilibrium and disequilibrium conditions. These five T–P–W profiles are: the 1D radiative–convective equilibrium profile (1D RCE) with a parametric $K_{ZZ}$ profile, the median global average profile (GA), the median east terminator average profile (ETA), the median west terminator average profile (WTA), and the median dayside average profile (DA). Figure~\ref{fig:TP_comp} shows these T–P profiles, while all 3D cases extended up to 10$^{-8}$~bar under isothermal conditions, along with the 1D self-consistent case where the top of the atmosphere is set to 10$^{-8}$~bar, are shown in Figure~\ref{fig:TP_comp_extended}. Figure \ref{fig:vmr_eq_term} illustrates the vertical profiles of the volume mixing ratios for six major opacity sources corresponding to the non-extended T–P profiles, whereas Figure \ref{fig:vmr_eq_term_extended} presents the same for the extended T–P profiles.
\begin{figure}
    \includegraphics[scale =0.32]{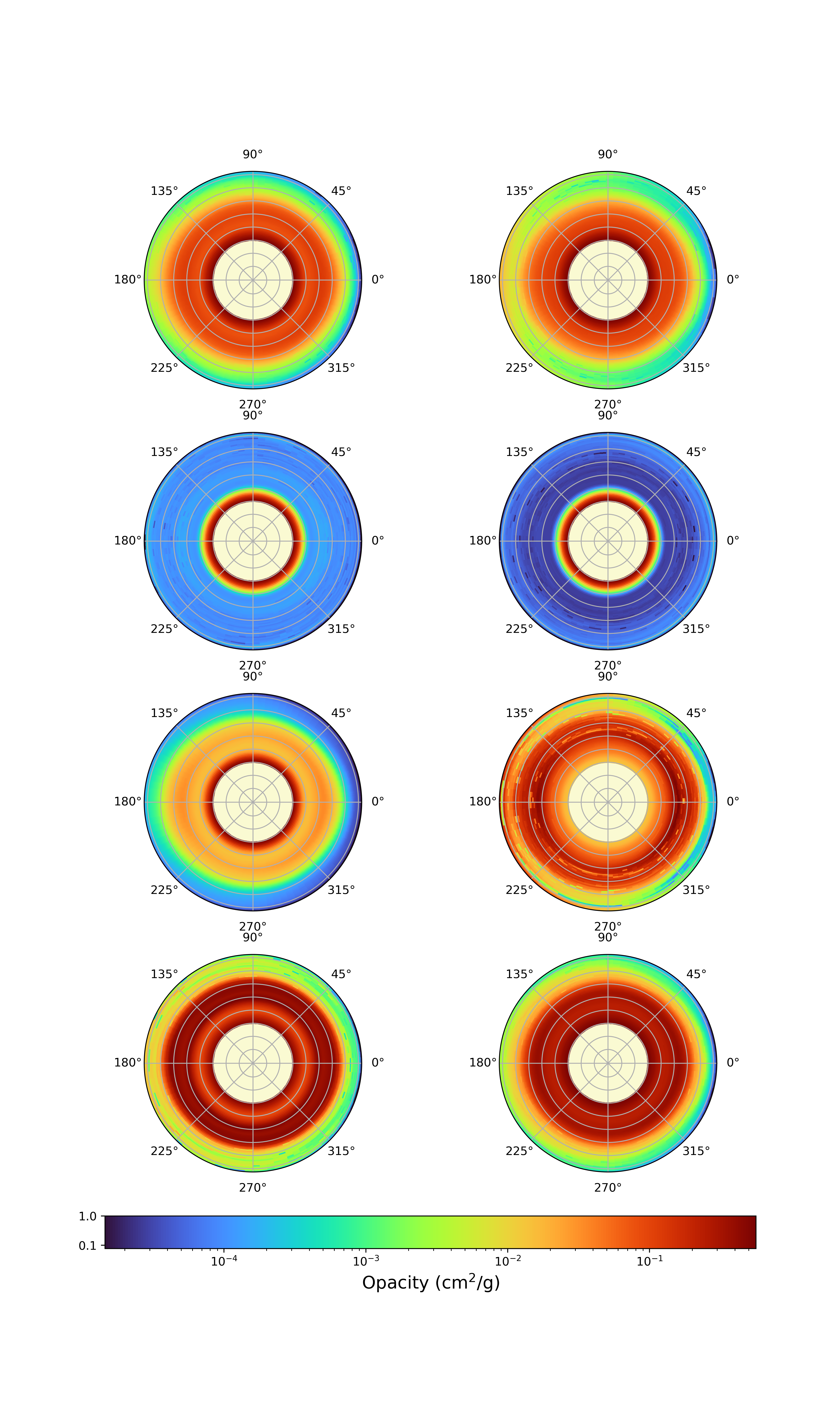}
    \caption{Polar maps of the total mass opacity used in the 3D radiative-transfer simulations. Each panel shows the spatial distribution of opacity across latitude and longitude, computed using equilibrium-chemistry opacities for the full 3D temperature structure. The eight subpanels correspond to opacity fields at eight wavelength points (1.28~$\mu$m, 1.54~$\mu$m, 2.24~$\mu$m, 2.28~$\mu$m, 3.2~$\mu$m, 4.32~$\mu$m, 7.71~$\mu$m, and 9.95~$\mu$m), illustrating how the wavelength and temperature dependence of line and continuum features maps onto spatial dependence in the atmosphere. The color scale shows the total mass opacity (cm$^{2}$\,g$^{-1}$) on a logarithmic scale.
}
    \label{fig:polar}
\end{figure}

\subsection*{Lower and Middle Atmosphere}
We observe that the VMR profiles of ETA and 1D RCE are largely similar due to their comparable T-P profiles, as shown in Figure \ref{fig:TP_comp}. The ETA and 1D RCE cases, having higher temperatures, exhibit higher abundances for all gases except HCN and CO in the upper parts of the atmosphere (< 0.1 bar). We also find that H$_2$O and CO exhibit different abundances between the 1D RCE and both ETA in the 3D model, as shown in Figures \ref{fig:vmr_eq_term}A and \ref{fig:vmr_eq_term}B. This difference is primarily observed in the lower regions of the atmosphere (100–10 bar), where equilibrium chemistry dominates. A similar trend is observed for the other four molecules in this region. 

For the disequilibrium case shown in Figure \ref{fig:vmr_eq_term}B, we found the following trends. In both the east terminator average (ETA) and west terminator average (WTA) 3D models, HCN abundances in the middle atmosphere (0.1--10$^{-5}$ bar) are regulated by a balance between formation through H$_2$ + CN $\rightarrow$ H + HCN and destruction via the three-body pathway HCN + H + M $\rightarrow$ H$_2$CN + M. These reactions proceed in both directions and dominate across both limbs, maintaining moderate HCN levels through reversible hydrogenation and radical cycling. In the 1D radiative-convective equilibrium (RCE) model, however, the reaction H$_2$CN + H $\rightarrow$ HCN + H$_2$ also contributes significantly to HCN formation. Nevertheless, HCN destruction via HCN + H + M $\rightarrow$ H$_2$CN + M occurs at a relatively higher rate in the 1D model than in the terminator models, resulting in a net decrease in HCN abundances at comparable pressures.

In both the ETA and WTA 3D models, NH$_3$ formation in the middle atmosphere (0.1--10$^{-5}$ bar) is dominated by NH$_2$ + H$_2$ $\rightarrow$ NH$_3$ + H. The reversible nature of this pathway, together with direct thermal decomposition of NH$_3$ into NH$_2$ + H, establishes a dynamic equilibrium that governs local NH$_3$ levels. An additional production channel, NH$_2$ + CH$_4$ $\rightarrow$ CH$_3$ + NH$_3$, is also active on both limbs, highlighting the role of hydrocarbon chemistry in nitrogen coupling. In the 1D RCE model, NH$_3$ formation is primarily driven by NH$_2$ + H$_2$ $\rightarrow$ NH$_3$ + H and the three-body reaction H + NH$_2$ + M $\rightarrow$ NH$_3$ + M, reflecting more efficient recombination under smoother vertical gradients and higher pressures.

In all three models, CH$_4$ formation in the middle atmosphere (0.1--10$^{-5}$ bar) is dominated by H + CH$_3$ + M $\rightarrow$ CH$_4$ + M, while destruction occurs via H + CH$_4$ $\rightarrow$ CH$_3$ + H$_2$ and its reverse. In the ETA and WTA models, CH$_4$ is also destroyed through oxidation by OH via OH + CH$_4$ $\rightarrow$ H$_2$O + CH$_3$, which is absent in the 1D RCE model. This difference leads to more efficient CH$_4$ depletion in the 3D models compared to the 1D case, where CH$_4$ remains relatively more stable due to fewer loss pathways.

At high pressures and temperatures, the chemical timescale becomes shorter than the eddy diffusion timescale, resulting in molecular abundances governed by equilibrium chemistry. This behavior is evident in Figure \ref{fig:vmr_eq_term}, where panels (A) and (B) show similar molecular abundances for the western and eastern terminators under both equilibrium and disequilibrium chemistry in the pressure range of 200 to 10 bar.

In both the dayside-average and global-terminator-average 3D models, HCN abundances in the middle atmosphere (0.1--10$^{-5}$ bar) are governed by the same bidirectional reactions as in the terminator models shown in subplot B, namely H$_2$ + CN $\rightleftharpoons$ H + HCN and HCN + H + M $\rightleftharpoons$ H$_2$CN + M. The global terminator model in subplot D shows chemical behavior nearly identical to that of the west terminator in subplot B, with all four reactions active and balanced. The 1D RCE model also retains the same pathways as in subplot B, indicating broadly consistent chemistry within this pressure range across all viewing geometries.

In subplot D, the middle-atmospheric chemistry of NH$_3$ closely parallels that in subplot B, with NH$_2$ + H$_2$ $\rightleftharpoons$ NH$_3$ + H dominating across all models. The dayside model includes an additional oxidation-driven loss via the reverse of NH$_3$ + OH $\rightarrow$ NH$_2$ + H$_2$O, similar to the west terminator in subplot B. The global terminator and 1D models show nearly identical behavior to the 1D RCE model in subplot B, relying on NH$_2$ + H$_2$ $\rightleftharpoons$ NH$_3$ + H and the three-body recombination reaction H + NH$_2$ + M $\rightarrow$ NH$_3$ + M for NH$_3$ formation.

In subplot D, CH$_4$ abundances in the middle atmosphere (0.1--10$^{-5}$ bar) are regulated by the same hydrogen-abstraction and recombination reactions as in subplot B, namely H + CH$_4$ $\rightleftharpoons$ CH$_3$ + H$_2$ and H + CH$_3$ + M $\rightarrow$ CH$_4$ + M. In the dayside model, CH$_4$ destruction is enhanced by OH + CH$_4$ $\rightarrow$ H$_2$O + CH$_3$, reflecting chemical behavior similar to that observed at the east terminator in subplot B. The global terminator model also includes this oxidation pathway, while the 1D model retains only the core abstraction and OH reactions, consistent with subplot B.

Similarly, Figure \ref{fig:vmr_eq_term} (C) presents three equilibrium VMR profiles derived from the T-P-W profiles of 1D RCE (hashed + dotted line), DA (solid line), and GA (hashed line), while Figure \ref{fig:vmr_eq_term} (D) shows the corresponding disequilibrium VMR profiles. The DA profile provides the best representation of the chemical composition that would be imprinted on the emission spectra\, as the emission spectra are generated on the dayside of the planet just before the secondary eclipse. We observe similar trends for DA and GA in the case of H$_2$O, CO, and CO$_2$, as seen in ETA and WTA (Figure \ref{fig:vmr_eq_term} A and C). All three models shown in Figure \ref{fig:vmr_eq_term} (D) share a common major formation pathway, characterized by the dissociation of CN by H$_2$: H$_2$ + CN $\rightarrow$ H + HCN. Despite having the same formation pathway, the reaction rates differ by several orders of magnitude across the three cases in the atmosphere above 1 bar, leading to significant variations in HCN abundance. 

Comparing the 1D RCE and DA profiles, we find similarities in their T-P profiles and equilibrium VMR profiles. The equilibrium abundances of gases show a direct correlation with temperature and pressure variations, which explains the similar VMR profiles observed in these two cases. Under disequilibrium conditions, the dayside profile exhibits behavior similar to that of the east and west terminator profiles due to small differences in the T-P and K$_{zz}$ profiles of these models.

\subsection*{Extended Upper Atmosphere}

In the upper atmosphere (pressures < 0.1 bar), where the transmission technique is most sensitive, the models predict an order-of-magnitude decrease in the abundances obtained from the 1D RCE model in Figure \ref{fig:vmr_eq_term_extended}A compared to those calculated for the terminator region (WTA), particularly for CH$_4$, HCN, and NH$_3$.

In the upper atmosphere ($<$10$^{-5}$ bar), photodissociation of HCN via HCN + h$\nu$ $\rightarrow$ H + CN becomes significant in both the ETA and WTA models, contributing to HCN depletion. However, concurrent recycling pathways such as H$_2$ + CN $\rightarrow$ H + HCN and H$_2$CN + H $\rightarrow$ HCN + H$_2$ help sustain HCN at low pressures, particularly on the east limb. In contrast, the 1D RCE model, which lacks explicit photochemical loss terms, maintains HCN primarily through hydrogenation and CN recombination, likely leading to an overestimation of HCN abundance in this region.

NH$_3$ is photodissociated through NH$_3$ + h$\nu$ $\rightarrow$ NH$_2$ + H in all three models. Despite this loss, NH$_3$ is partially replenished through NH$_2$ + H$_2$ $\rightarrow$ NH$_3$ + H, though the efficiency of this recycling varies. On the west limb, minor contributions arise from the reverse reaction NH$_3$ + CH $\rightarrow$ NH$_2$ + CH$_2$, while in the 1D model, the reverse of NH$_3$ + OH $\rightarrow$ NH$_2$ + H$_2$O also plays a role. The net result is a decline in NH$_3$ abundance with altitude in all models, although the magnitude of depletion differs due to the interplay between photochemistry and vertical mixing. The substantial variation in NH$_3$ abundances can be attributed to an order-of-magnitude difference in formation rates, likely caused by a high $K_{ZZ}$ value, which leads to shorter kinetic and longer chemical timescales, thereby reducing reaction efficiencies.

CH$_4$ undergoes photodissociation in the 3D models through CH$_4$ + h$\nu$ $\rightarrow$ CH$_2$ + H$_2$, while also participating in reversible hydrogen abstraction and recombination (H + CH$_3$ + M $\rightarrow$ CH$_4$ + M). These processes counteract each other, with photolysis driving CH$_4$ loss and radical chemistry partially restoring it. In contrast, the 1D model lacks photodissociation and maintains higher CH$_4$ levels, with destruction occurring mainly through H + CH$_4$ $\rightarrow$ CH$_3$ + H$_2$ and oxidation by OH. This leads to an overprediction of CH$_4$ abundance at low pressures in the 1D framework.

Photodissociation of HCN via HCN + h$\nu$ $\rightarrow$ CN + H is enhanced in the dayside model, leading to enhanced HCN depletion compared to both the global terminator and 1D models. The latter two rely solely on the reversible H$_2$ + CN $\rightleftharpoons$ H + HCN and HCN + H + M $\rightleftharpoons$ H$_2$CN + M reactions, as in subplot B, resulting in higher HCN abundances at low pressures. The presence of photolytic destruction in the dayside case, partially offset by H$_2$CN + H $\rightarrow$ HCN + H$_2$, introduces a key divergence from the more shielded regions represented in subplot B.

\subsection{Modelled synthetic transmission and emission spectrum}

\subsubsection{Non-extended Cases}

\begin{figure*}
    \centering
    \includegraphics[scale =0.45]{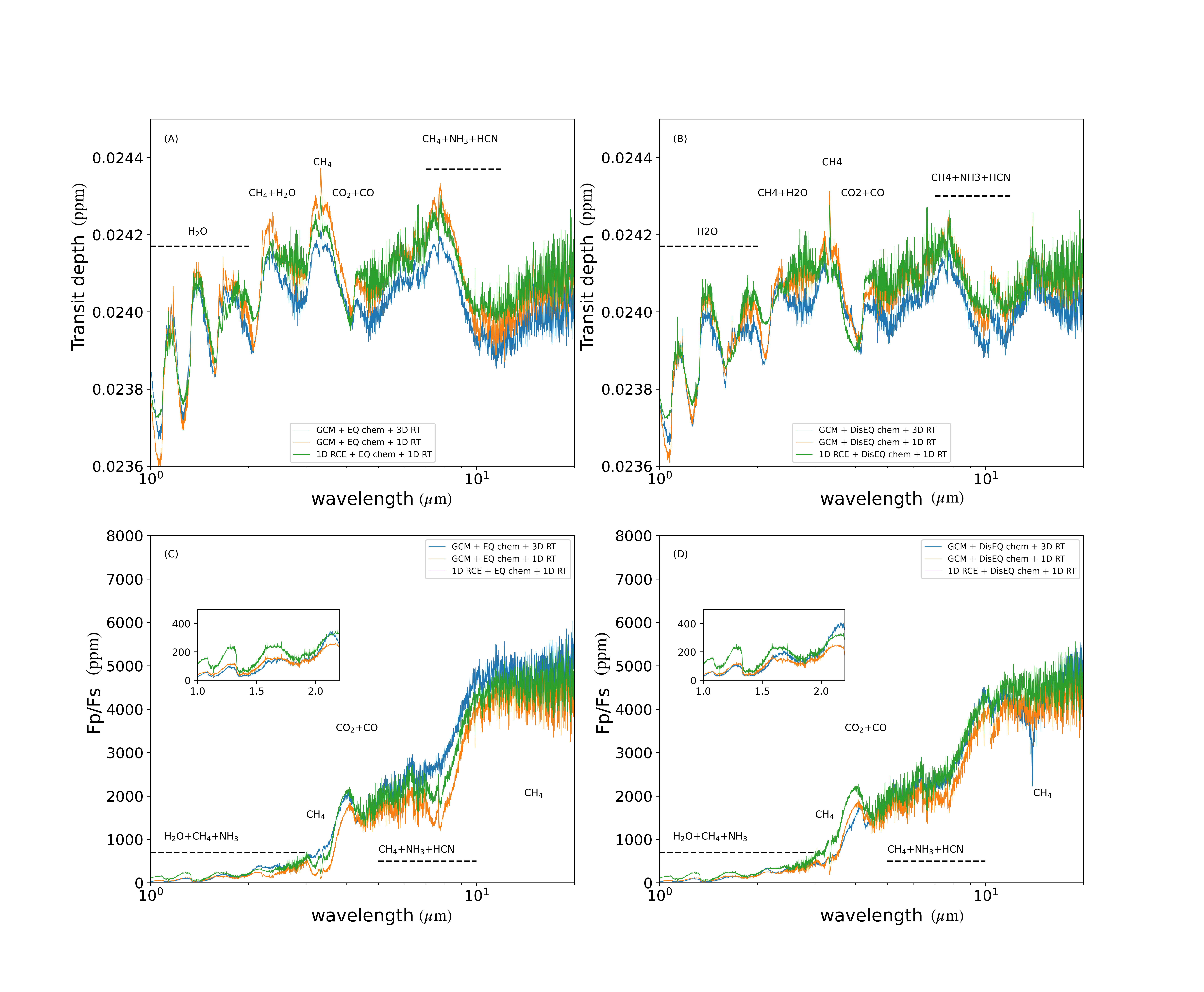}
    \caption{Subfigures (A) and (B) show the transmission spectrum using equilibrium and disequilibrium chemistry, respectively, from the three different cases compared in our study: complete 1D modeling (green), 3D GCM (GA) with 1D chemical profiling and 1D radiative transfer (orange), and 3D GCM (ETA and WTA average) with 1D chemical profiling and 3D radiative transfer (blue) using the T–P profiles shown in Figure \ref{fig:TP_comp}. Subfigures (C) and (D) represent the emission spectrum using equilibrium and disequilibrium chemistry, respectively, with a color scheme similar to (A) and (B).}
    \label{fig:eq_trans}
\end{figure*}
\begin{figure*}
    \centering
    \includegraphics[scale =0.45]{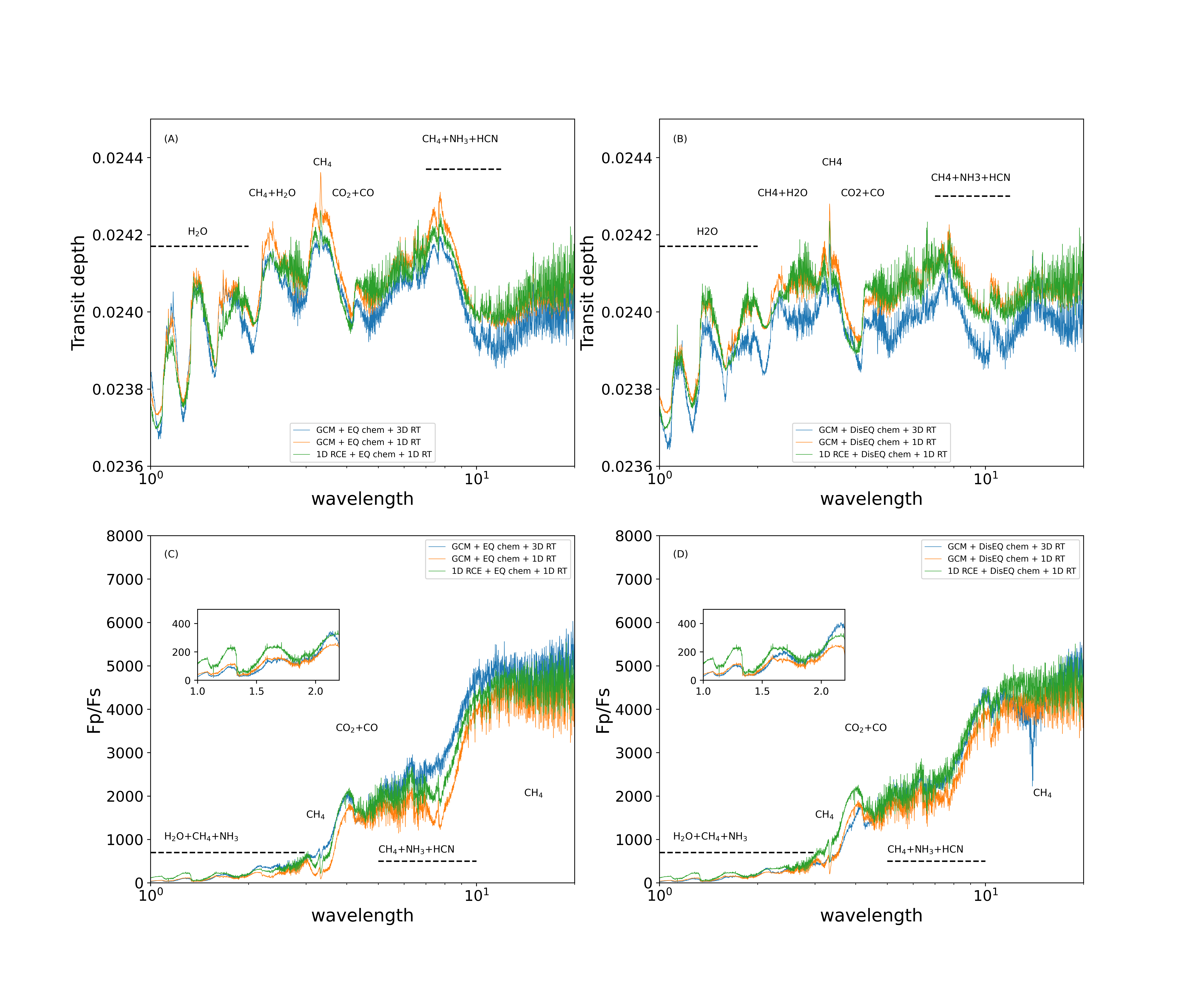}
    \caption{Subfigures (A) and (B) show the transmission spectra obtained using equilibrium and disequilibrium chemistry, respectively, while Subfigures (C) and (D) present the corresponding emission spectra for equilibrium and disequilibrium chemistry, computed using the T–P profiles shown in Figure \ref{fig:TP_comp_extended}. The color scheme follows that of (A) and (B), comparing the extended 1D RCE + 1D RT and 3D GCM + 1D RT spectra against the non-extended 3D GCM + 3D RT spectra for reference. Due to the limitation described in Section~\ref{subsec:limit}, the full 3D GCM + 3D RT configuration is not available for the extended setup.}
    \label{fig:eq_trans_extended}
\end{figure*}

We present four comparison cases in Figure \ref{fig:eq_trans}. Figure \ref{fig:eq_trans}A shows the transmission spectra computed under chemical equilibrium for (i) a full 1D model, (ii) a 3D GCM (GA) with 1D chemical profiling and 1D radiative transfer, and (iii) a 3D GCM (ETA and WTA averages) with 1D chemical profiling and 3D radiative transfer and (iv) a 3D GCM (DA) with 1D chemical profiling and 3D radiative transfer, using the T–P profiles shown in Figure \ref{fig:TP_comp}. Figure \ref{fig:eq_trans}B presents the same cases but with disequilibrium chemistry. Similarly, Figure \ref{fig:eq_trans}C shows emission spectra using equilibrium chemistry for (i) a complete 1D model, (ii) a 3D GCM (GA) with 1D chemical profiling and 1D radiative transfer, and (iii) a 3D GCM (DA) with 1D chemical profiling and 3D radiative transfer. Figure \ref{fig:eq_trans}D presents the same cases but with disequilibrium chemistry. Differences in molecular abundances and the radiative transfer method are the two primary factors contributing to the discrepancies between our 1D and 3D models. We discuss the effects and variations in molecular abundances in Section \ref{subsec:VMR}. For radiative transfer, we use \texttt{petitRADTRANS} for the 1D model and \texttt{gCMCRT} for the 3D model. Due to differences in their working dimensionality, these models take different temperature-pressure distributions as input. For the 3D simulation, even when using the same volume mixing ratios as the 1D radiative transfer simulation, inhomogeneous opacity is observed, as shown in Figure \ref{fig:polar}. This demonstrates how, for the same VMR, variations in opacity in the 3D simulations lead to differences in the transmission and emission spectra compared to the 1D radiative transfer model. The opacity plots indicate higher opacity toward the nightside, which corresponds to lower temperatures and pressures, whereas lower opacity is observed on the dayside, where temperatures and pressures are higher. These differences result in an overall decrease in transit depth for the 3D model, as discussed in Section \ref{subsec:ts}.

\begin{table*}
\caption{This table presents SNR data for four cases: transmission and emission spectra in both equilibrium and disequilibrium conditions. The first two columns represent the wavelength region of the eight major spectral features and the corresponding molecule. Columns 3 and 4 show the signal strength in the 1D and 3D models (transit depth minus baseline transit depth / emission flux). Columns 5 and 6 present the absolute difference between these values and the SNR achieved with three transits of observation.In the non-extended cases, the “Difference (Non-extended)” column represents the difference between the full 1D RCE + 1D chemistry + 1D radiative transfer model and the 3D GCM + 1D chemistry + 3D radiative transfer model, whereas “Difference (Extended)” represents the difference between 1D RCE + 1D chemistry + 1D radiative transfer model (extended) and the 3D GCM + 1D chemistry + 3D radiative transfer model (non extended). The SNR for the extended cases is given in brackets next to the original SNR values, which correspond to the non-extended cases.
}
\label{tab:restab}

\begin{tabular}{lllllll}
\hline
\multicolumn{7}{l}{\textbf{Transmission Spectra (Disequilibrium Chemistry)}} \\ \hline \hline
Molecule        & Wavelength ($\mu$m) & 1D Signal  & 3D Signal  & Difference & Difference  & SNR  \\ 
& ($\mu$m) & (ppm) & (ppm) & (Non-extended, ppm) & (Extended, ppm) & (3 Transits) \\ 
\hline
H$_2$O          & 1.1-1.35            & 294 (148) & 222 & 72 & 74 & 5.95 (6.11) \\
H$_2$O          & 1.35-1.6            & 384 (342) & 303 & 81 & 39 & 4.25 (2.05) \\
H$_2$O          & 1.7-2.25            & 270 (261) & 161 & 109 & 100 & 5.32 (4.88) \\
H$_2$O + CH$_4$ & 2.15-3              & 514 (488) & 364 & 150 & 124 & 7.04 (5.82) \\
CH$_4$          & 3.32                & 577 (535) & 491 & 88 & 44 & 3.30 (1.65) \\
CO + CO$_2$     & 3.7-5.1             & 401 (210) & 276 & 125 & 66 & 5.07 (2.68) \\
CH$_4$          & 7.4-7.74            & 564 (448) & 464 & 100 & 16 & 3.15 (0.50) \\
NH$_3$          & 9.86-12             & 405 (345) & 280 & 125 & 65 & 1.66 (0.87) \\ \hline \\ \hline

\multicolumn{7}{l}{\textbf{Transmission Spectra (Equilibrium Chemistry)}} \\ \hline \hline
Molecule        & Wavelength ($\mu$m) & 1D Signal  & 3D Signal  & Difference & Difference  & SNR  \\ 
& ($\mu$m) & (ppm) & (ppm) & (Non-extended, ppm) & (Extended, ppm) & (3 Transits) \\ 
\hline
H$_2$O          & 1.1-1.35            & 237 (186) & 330 & 93 & 144 & 6.10 (9.45) \\
H$_2$O          & 1.35-1.6            & 207 (202) & 147 & 60 & 55 & 6.21 (5.69) \\
H$_2$O          & 1.7-2.25            & 310 (274) & 207 & 103 & 67 & 5.70 (3.71) \\
H$_2$O + CH$_4$ & 2.15-3              & 520 (494) & 383 & 137 & 111 & 7.68 (6.22) \\
CH$_4$          & 3.32                & 599 (563) & 509 & 90 & 54 & 3.17 (1.90) \\
CO + CO$_2$     & 3.7-5.1             & 461 (375) & 323 & 138 & 52 & 7.03 (2.65) \\
CH$_4$          & 7.4-7.74            & 601 (555) & 494 & 107 & 61 & 2.55 (1.45) \\
NH$_3$          & 9.86-12             & 274 (341) & 153 & 121 & 188 & 2.04 (3.17) \\ \hline \\ \hline

\multicolumn{7}{l}{\textbf{Emission Spectra (Disequilibrium Chemistry)}} \\ \hline \hline
Molecule        & Wavelength ($\mu$m) & 1D Signal  & 3D Signal  & Difference & Difference  & SNR  \\ 
& ($\mu$m) & (ppm) & (ppm) & (Non-extended, ppm) & (Extended, ppm) & (3 Transits) \\ 
\hline
H$_2$O          & 1.1-1.35            & 253 (233) & 105 & 148 & 128 & 14.08 (12.18) \\
H$_2$O          & 1.35-1.6            & 249 (243) & 131 & 118 & 112 & 7.11 (6.75) \\
H$_2$O          & 1.7-2.25            & 258 (336) & 172 & 86 & 164 & 7.92 (15.10) \\
H$_2$O + CH$_4$ & 2.15-3              & 645 (638) & 424 & 221 & 214 & 8.30 (8.03) \\
CH$_4$          & 3.32                & 216 (1004) & 501 & 285 & 503 & 16.12 (28.46) \\
CO + CO$_2$     & 3.7-5.1             & 1945 (2082) & 1200 & 545 & 882 & 15.98 (25.87) \\
CH$_4$          & 7.4-7.74            & 1865 (2759) & 2552 & 687 & 207 & 36.93 (11.13) \\
NH$_3$          & 9.86-12             & 4850 (4762) & 4043 & 807 & 719 & 12.77 (11.38) \\ \hline \\ \hline

\multicolumn{7}{l}{\textbf{Emission Spectra (Equilibrium Chemistry)}} \\ \hline \hline
Molecule        & Wavelength ($\mu$m) & 1D Signal  & 3D Signal  & Difference & Difference  & SNR  \\ 
& ($\mu$m) & (ppm) & (ppm) & (Non-extended, ppm) & (Extended, ppm) & (3 Transits) \\ 
\hline
H$_2$O          & 1.1-1.35            & 239 (238) & 103 & 136 & 135 & 15.43 (15.31) \\
H$_2$O          & 1.35-1.6            & 204 (201) & 117 & 87 & 84 & 9.80 (9.46) \\
H$_2$O          & 1.7-2.25            & 267 (266) & 379 & 112 & 113 & 5.90 (5.95) \\
H$_2$O + CH$_4$ & 2.15-3              & 251 (270) & 259 & 108 & 11 & 10.54 (1.07) \\
CH$_4$          & 3.32                & 297 (217) & 668 & 371 & 451 & 12.41 (15.08) \\
CO + CO$_2$     & 3.7-5.1             & 1983 (1953) & 1541 & 442 & 412 & 34.52 (32.15) \\
CH$_4$          & 7.4-7.74            & 1865 (1889) & 2868 & 1004 & 979 & 23.35 (22.77) \\
NH$_3$          & 9.86-12             & 4884 (3794) & 5485 & 601 & 1691 & 8.01 (22.53) \\ \hline
\end{tabular}

\end{table*}

\subsubsection{Transmission Spectra}\label{subsec:ts}
We modeled synthetic transmission spectra for three cases, as defined in Section \ref{sec:trans_model}, using both equilibrium and disequilibrium chemistry. Figures \ref{fig:eq_trans} (A) and (B) compare the transmission spectra of the three modeled cases.

In the case of equilibrium chemistry, shown in Figure \ref{fig:eq_trans} (A), the differences between the orange and green curves arise solely from variations in the T-P profiles, which affect the 1D equilibrium molecular abundances. In contrast, the differences between the blue and orange curves result from the use of different radiative transfer methods. The 3D radiative transfer (RT) technique underestimates spectral features compared to the 1D RT method. We observe that H$_2$O, CH$_4$, CO$_2$, and CO exhibit the most distinguishable spectral differences between the 1D and 3D models, with variations exceeding 100 ppm. A more detailed quantitative assessment of the wavelength-dependent signal differences is provided in Table \ref{tab:restab}.

Figure \ref{fig:eq_trans} (B) presents the synthetic transmission spectra for disequilibrium chemistry. We find that the difference in transit depth between the 1D and 3D models decreases for all molecules except for the combined H$_2$O + CH$_4$ feature in the 2.15–3 $\mu$m range.

Furthermore, we identify significant differences between the equilibrium and disequilibrium spectra. All major H$_2$O features exhibit greater transit depths in equilibrium models compared to disequilibrium models, with differences exceeding 50 ppm in the 1D models and 100 ppm in the 3D models. Equilibrium chemistry also predicts a higher CH$_4$ abundance than disequilibrium conditions, as CH$_4$ readily photolyzes into CH$_3$ and H a process not accounted for in equilibrium chemistry.

\subsubsection{Emission Spectra}

In our study, the median dayside-averaged profile used for 3D RT is the hottest, followed by the 1D RCE profile, while the global median profile is the coldest. Figures \ref{fig:eq_trans} (C) and (D) illustrate these temperature variations. We also find that differences between the emission spectra of the 1D and 3D models are highly wavelength-dependent, with larger deviations at longer wavelengths due to the higher emission flux from the planet at those wavelengths.

Figure \ref{fig:eq_trans} (C) shows the synthetic emission spectra under equilibrium chemistry for HD 189733b. In this case, we observe significant differences between the 1D and 3D models (371–1004 ppm), primarily attributed to CH$_4$, CO, CO$_2$, and NH$_3$. Unlike the transmission spectra, the 1D and 3D emission spectra exhibit no notable differences in H$_2$O signals.

Figure \ref{fig:eq_trans} (D) presents the synthetic emission spectra under disequilibrium chemistry. The disparity between the 1D and 3D models decreases under disequilibrium conditions, similar to what we observed for the transmission spectra. As in the equilibrium case, differences of 285–807 ppm arise, mainly due to the same opacity sources (CH$_4$, CO, CO$_2$, and NH$_3$). CH$_4$ is the most significantly affected opacity source under disequilibrium conditions, as evidenced by the sharp decrease in its 7.4–7.74 $\mu$m and 14 $\mu$m features in Figure \ref{fig:eq_trans} (D). The impact of CH$_4$ photodissociation (Section \ref{subsec:VMR}) on signal strength is more pronounced in the 3D models, indicating that they are more sensitive to the treatment of chemistry compared to 1D models.

\subsubsection{Extended Cases and Limitations}\label{subsec:limit}

A key limitation of our extended cases arises from the numerical pressure ceiling in THOR, which typically reaches up to $10^{-4}$~bar \citep{THOR+HELIOS}. For photochemical modeling, we extrapolated these T–P–W profiles isothermally to $10^{-8}$~bar to capture photochemical effects. However, since \texttt{gCMCRT} is built on a constant-height grid, extending the grid to $10^{-8}$~bar on the dayside causes the corresponding nightside levels to fall below $10^{-10}$~bar, resulting in numerical instabilities. Consequently, we generate spectra for the extended cases only with the 1D radiative transfer model to examine how the extended atmospheres compare with the non-extended cases and to assess the impact of photochemistry. The extended atmospheric spectra are shown in Figure~\ref{fig:eq_trans_extended} for the 1D RCE + 1D chemistry + 1D RT and 3D GCM + 1D chemistry + 1D RT configurations, alongside the non-extended spectra (3D GCM + 1D chemistry + 3D RT), enabling direct comparison between the extended and non-extended cases and illustrating the signal differences listed in Table~\ref{tab:restab}. Comparing the 1D RCE + 1D chemistry + 1D RT with the 3D GCM + 1D chemistry + 1D RT reflects only the differences in the T–P profiles, rather than in the modeling techniques, which are the main focus of this study. Therefore, in Table \ref{tab:restab}, the extended difference case represents the difference between the extended 1D RCE + 1D chemistry + 1D RT and the non-extended 3D GCM + 1D chemistry + 3D RT. This comparison highlights the impact of the lower-pressure upper atmosphere.

\subsubsection{Transmission Spectra}

In transmission, the extended 1D–3D comparison shows that the largest contrasts occur in water-dominated regions and in the blended H$_2$O+CH$_4$ band. Under disequilibrium chemistry, the inclusion of the extended upper atmosphere produces only small changes in most features, with differences varying by less than 20~ppm for H$_2$O bands and decreasing substantially for CH$_4$ and CO/CO$_2$ regions, in some cases by nearly half. In the equilibrium case, the effect of the extended atmosphere is more pronounced: water features and the NH$_3$ window show stronger enhancements, with extended differences increasing by 40–80~ppm, whereas CH$_4$ and CO/CO$_2$ exhibit a substantial reduction in model differences.

\subsubsection{Emission Spectra}

In emission, the extended 1D–3D comparison shows that the largest spectral differences occur in the CH$_4$, CO/CO$_2$, and NH$_3$ bands, while H$_2$O features exhibit comparatively smaller variations. Under disequilibrium chemistry, extending the atmosphere enhances the contrast in CH$_4$ and CO/CO$_2$ regions by several hundred ppm. In contrast, H$_2$O bands vary only modestly, typically within 20–40~ppm, indicating limited sensitivity in the near-infrared features. 

In the equilibrium case, the overall behavior remains similar but with smaller enhancements in the CH$_4$ and CO/CO$_2$ features and slightly larger increases in the H$_2$O and NH$_3$ windows. These results suggest that disequilibrium chemistry amplifies wavelength-dependent contrasts between the 1D and 3D frameworks in the extended cases, whereas equilibrium chemistry produces a more uniform offset across most bands, except for CH$_4$ and NH$_3$.

\subsection{Noise simulation and Signal-to-Noise ratio}

\begin{figure*}
    \centering
    \includegraphics[scale = 0.8]{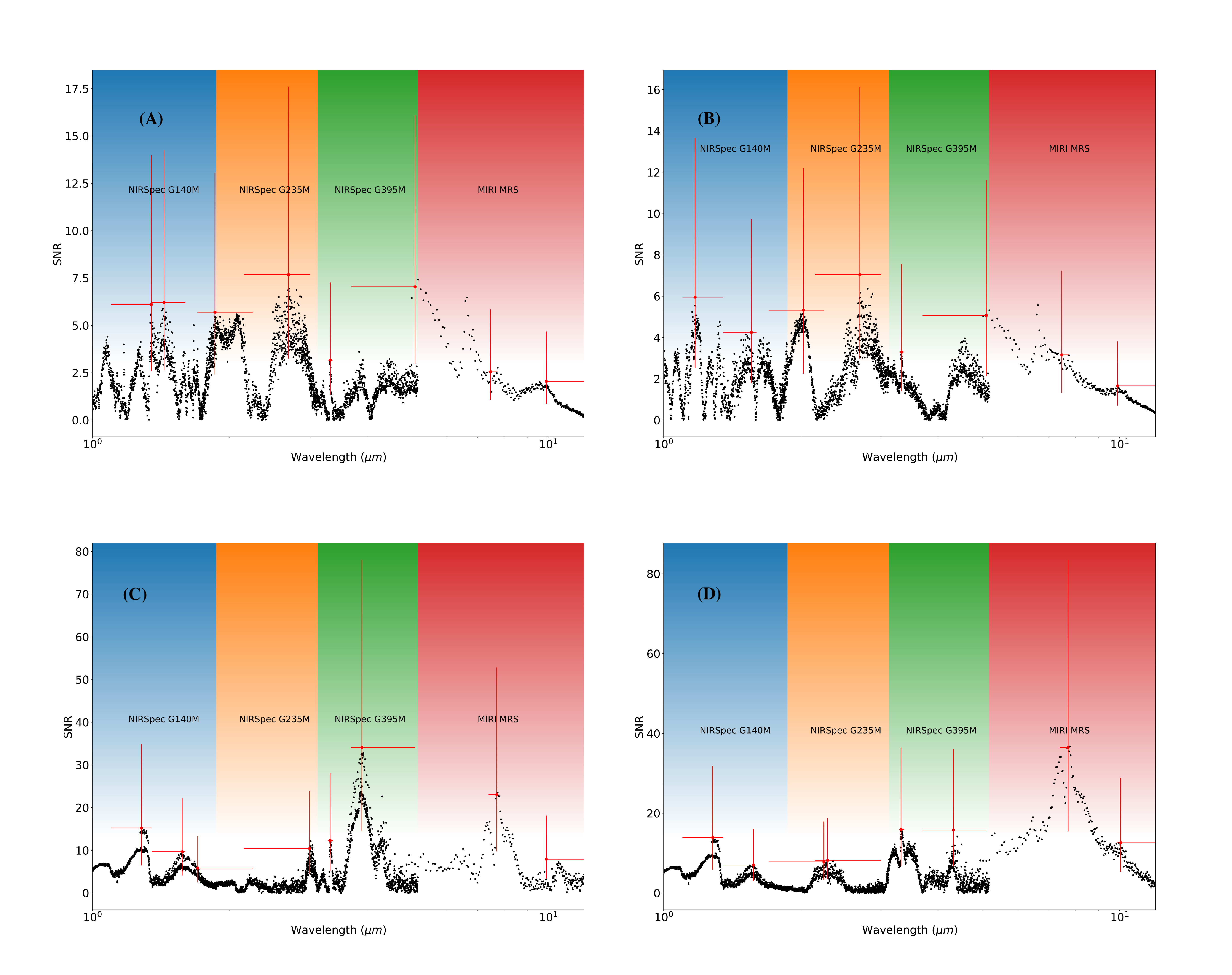}
    \caption{The black dotted curve shows the SNR for the differences between the 1D and 3D models over three transits for the non-extended cases, as listed in Table~\ref{tab:restab}. The red scatter plot shows the SNR achieved over three transits for the eight major features shown in Table \ref{tab:restab}, with upper and lower error bars corresponding to five and one transits, respectively. The figure represents four cases of simulation: transmission with equilibrium chemistry (A), transmission with disequilibrium chemistry (B), emission with equilibrium chemistry (C), and emission with disequilibrium chemistry (D).}
    \label{fig:diff_eqt}
\end{figure*}

We used our models to generate noise simulations and computed the signal-to-noise ratio (SNR) for the differences between the spectra of the non-extended and extended cases, as shown in Table~\ref{tab:restab}. These SNRs were evaluated over all wavelengths for observations consisting of three transits. Additionally, we calculated the SNR for one and five transits, as shown in Figure~\ref{fig:diff_eqt}, where error bars for the non-extended cases are included for illustration.

To better understand variations in the major spectral features arising from different opacity sources, we identified eight key features in the 1–12~$\mu$m range. These correspond to the following wavelength intervals and associated molecular contributions: 1.1–1.35~$\mu$m (H$_2$O), 1.35–1.6~$\mu$m (H$_2$O), 1.7–2.25~$\mu$m (H$_2$O), 2.15–3~$\mu$m (H$_2$O + CH$_4$), 3.32~$\mu$m (CH$_4$), 3.7–5.1~$\mu$m (CO + CO$_2$), 7.4–7.74~$\mu$m (CH$_4$), and 9.86–12~$\mu$m (NH$_3$).

Figure~\ref{fig:diff_eqt} (A) shows the achieved signal-to-noise ratio (SNR) as a function of wavelength for transmission under equilibrium chemistry. The SNR calculations indicate that H$_2$O (SNR = 5.7–7.68), CO$_2$, and CO (SNR = 7.03) are the molecules capable of distinguishing between the 1D and 3D models, with an SNR greater than 5 in three transits (Table~\ref{tab:restab}). The figure also shows that increasing the number of observations from 3 to 5 transits enhances the detectability of CH$_4$, making it an important distinguishing marker.

As anticipated from the differences between the 1D and 3D models discussed in Section~\ref{subsec:ts}, we observe a decrease in SNR for disequilibrium chemistry, except for CH$_4$, as shown in Figure~\ref{fig:diff_eqt} (B). Additionally, we find that SNR increases as signal strength decreases for the same NIRSpec G140M grism. This trend arises due to the wavelength-dependent efficiency of the grism, defined by its response function. For example, the NIRSpec G140M grism covers the 0.97–1.89~$\mu$m range, with peak efficiency around 1.2~$\mu$m.

A similar analysis for the equilibrium emission spectrum is shown in Figure~\ref{fig:diff_eqt} (C). The SNR achieved in distinguishing the 1D and 3D (DA) models in three transits of emission is more than twice that obtained for transmission. Unlike the transmission spectra, we find that CH$_4$ (SNR = 12.41–23.35), along with H$_2$O (SNR = 5.9–15.43), CO$_2$, and CO (SNR = 34.52), can also serve as key tracers for differentiating between the 1D and 3D models. For the emission spectrum under disequilibrium chemistry, our analysis reveals trends similar to those observed between equilibrium and disequilibrium transmission spectra.

A combined analysis highlights the 3.7–5.1~$\mu$m feature, attributed to CO and CO$_2$, as the most critical spectral region for distinguishing between 1D and 3D models in all cases examined. Additionally, we identify case-specific distinguishing features: H$_2$O for transmission spectra and CH$_4$ for emission spectra, both of which provide insights into the planet's three-dimensional atmospheric structure.

We repeated the comparison using the extended atmospheric spectra (Fig.~\ref{fig:eq_trans_extended}) and Table~\ref{tab:restab}. In transmission, extending the 1D model leads to moderate but structured variations in SNR across wavelengths and chemical regimes. Under disequilibrium chemistry, most bands show a decrease in SNR, with reductions of nearly 50\% in the CH$_4$ (3.3~$\mu$m) and CO/CO$_2$ (3.7–5.1~$\mu$m) regions, and smaller declines of 10–30\% in the H$_2$O bands. Despite these reductions, the H$_2$O and blended H$_2$O+CH$_4$ regions remain the strongest contributors, retaining SNR values above 5 even after extension. In contrast, equilibrium chemistry shows the opposite trend: SNRs increase in the long-wavelength H$_2$O and NH$_3$ windows by up to a factor of 1.5, reflecting the enhanced sensitivity of these bands to upper-atmosphere temperature gradients. CH$_4$ and CO/CO$_2$ features show moderate decreases in the extended case, indicating that the radiative structure of the lower atmosphere dominates their contribution. Overall, extending the model redistributes SNR between the near- and mid-infrared but preserves the hierarchy of dominant features, with equilibrium chemistry amplifying long-wavelength contrasts and disequilibrium chemistry reducing them.

In emission, the extended models show a wider range of SNR variations. Under disequilibrium chemistry, the CH$_4$ (3.3~$\mu$m) and CO/CO$_2$ (3.7–5.1~$\mu$m) bands show the largest SNR increases, by roughly 70–90\% relative to the non-extended case, while H$_2$O features near 1–2~$\mu$m change only slightly, remaining within 10–15\%. At longer wavelengths, NH$_3$ (9.9–12~$\mu$m) shows a small SNR decrease. Under equilibrium chemistry, SNR variations are more balanced: CH$_4$ and CO/CO$_2$ remain strong (SNR~$\sim$30–40) but decrease slightly, while NH$_3$ and H$_2$O bands show modest enhancements.

\section{Summary and Conclusion} \label{sec:conc}

We modeled the atmosphere of HD 189733b to quantify the differences between 1D and 3D models, analyzing the required observational time to distinguish between these models and understand the 3D features of this planet. Our key findings and conclusions are summarized below:

\begin{itemize}
    \item Our GCM models suggest inefficient heat redistribution in the dense lower atmosphere, resulting in homogeneous temperatures, while high wind velocities in the upper atmosphere drive efficient heat transport, creating notable variations. An eastward hotspot shift of 16$^{\circ} \pm 4^{\circ}$, consistent with Spitzer and HST observations, highlights the influence of atmospheric winds driven by planetary rotation.
    
    \item The terminator region and median dayside-averaged T-P profiles differ significantly from the 1D T-P profile, highlighting the critical need for 3D modeling over 1D modeling, as transmission and emission spectra are sensitive to these regions rather than median global averages.
    
    \item The comparison of molecular abundances between 3D GCM-derived terminator/dayside profiles and 1D radiative-convective equilibrium models reveals significant differences. While 1D models predict higher abundances for most gases, key species like HCN, NH$_3$, and CH$_4$ show marked discrepancies due to differences in reaction pathways and rates.
    
    \item Synthetic transmission and emission spectra modeled for the 1D and 3D cases with equilibrium and disequilibrium chemistry reveal notable differences in key opacity sources such as H$_2$O, CH$_4$, CO, CO$_2$, and NH$_3$, where the T–P profiles extend to higher-pressure regions (10$^{-4}$~bar for 3D cases and 10$^{-6}$~bar for 1D cases, defined as non-extended cases). In these cases, equilibrium models predict stronger CH$_4$ and H$_2$O features, whereas disequilibrium models show reduced transit depths due to photochemical effects. Differences in the emission spectra between the 1D and 3D models are wavelength-dependent, with the 3D models showing greater sensitivity to chemical processes, particularly in the CH$_4$ features.

    \item For the non-extended cases, key spectral features in the 1–12~$\mu$m range, including H$_2$O, CO$_2$, CO, and CH$_4$, serve as primary markers for distinguishing between models, with SNR~$>$~5 achieved in most instances over three transits under equilibrium chemistry. Emission spectra exhibit higher SNRs than transmission spectra, with CH$_4$ and the combined CO+CO$_2$ bands emerging as critical indicators. Although disequilibrium chemistry generally reduces SNR values, CH$_4$ remains an exception. The 3.7–5.1~$\mu$m range, dominated by CO and CO$_2$, consistently emerges as the most significant region for model differentiation across all scenarios. 

    \item For the non-extended cases, we show that an observation time of three transits, equivalent to 11.4~hours (5.4~hours for the transit and 6~hours for the baseline), is required to achieve an SNR~$>$~5 in transmission, while approximately one transit, equivalent to 3.8~hours (1.8~hours for the transit and 2~hours for the baseline), is sufficient to reach the same SNR in emission. These results provide an indirect estimate of the minimum observational time needed to distinguish between 1D and 3D atmospheric models for hot, irradiated gaseous exoplanets such as HD~189733b.

    \item For the extended cases, where lower pressures up to 10$^{-8}$~bar modify the SNR in a wavelength- and chemistry-dependent manner, the near-infrared CH$_4$ and CO/CO$_2$ emission features, along with the long-wavelength H$_2$O and NH$_3$ transmission bands, are enhanced under equilibrium chemistry. In contrast, disequilibrium chemistry reduces the overall contrast and smooths SNR variations across the spectrum.
  
\end{itemize}

\medskip
\section*{acknowledgment}
L.M. acknowledges funding support from the DAE through the NISER project RNI 4011.  L.M. also acknowledges helpful discussions on 3D GCMs with Nicolas Iro during the Cloud Academy 3 workshop at the Les Houches Advanced School of Physics in the French Alps. R.A. acknowledges valuable conversations with Elspeth K. H. Lee on gCMCRT. We thank the anonymous referee for constructive comments that helped improve the manuscript.

\section*{Data Availability}
All the codes used in this study, such as FASTCHEM, VULCAN, THOR, HELIOS, petitRADTRANS, and gCMCRT, are publicly available. All simulated data created in this study will be shared upon reasonable request to the corresponding author.

\bibliographystyle{mnras}
\bibliography{example} 

@ARTICLE{madhu2019,
       author = {{Madhusudhan}, Nikku},
        title = "{Exoplanetary Atmospheres: Key Insights, Challenges, and Prospects}",
      journal = {\araa},
         year = 2019,
        month = aug,
       volume = {57},
        pages = {617-663},
          doi = {10.1146/annurev-astro-081817-051846},
archivePrefix = {arXiv},
       eprint = {1904.03190},
 primaryClass = {astro-ph.EP},
       adsurl = {https://ui.adsabs.harvard.edu/abs/2019ARA&A..57..617M}
}

@article{guillot2010,
  title={On the radiative equilibrium of irradiated planetary atmospheres},
  author={Guillot, Tristan},
  journal={Astronomy \& Astrophysics},
  volume={520},
  pages={A27},
  year={2010},
  publisher={EDP Sciences}
}

@ARTICLE{ers23,
       author = {{JWST Transiting Exoplanet Community Early Release Science Team} and {Ahrer}, Eva-Maria and {Alderson}, Lili and {Batalha}, Natalie M. and {Batalha}, Natasha E. and {Bean}, Jacob L. and {Beatty}, Thomas G. and {Bell}, Taylor J. and {Benneke}, Bj{\"o}rn and {Berta-Thompson}, Zachory K. and {Carter}, Aarynn L. and {Crossfield}, Ian J.~M. and {Espinoza}, N{\'e}stor and {Feinstein}, Adina D. and {Fortney}, Jonathan J. and {Gibson}, Neale P. and {Goyal}, Jayesh M. and {Kempton}, Eliza M. -R. and {Kirk}, James and {Kreidberg}, Laura and {L{\'o}pez-Morales}, Mercedes and {Line}, Michael R. and {Lothringer}, Joshua D. and {Moran}, Sarah E. and {Mukherjee}, Sagnick and {Ohno}, Kazumasa and {Parmentier}, Vivien and {Piaulet}, Caroline and {Rustamkulov}, Zafar and {Schlawin}, Everett and {Sing}, David K. and {Stevenson}, Kevin B. and {Wakeford}, Hannah R. and {Allen}, Natalie H. and {Birkmann}, Stephan M. and {Brande}, Jonathan and {Crouzet}, Nicolas and {Cubillos}, Patricio E. and {Damiano}, Mario and {D{\'e}sert}, Jean-Michel and {Gao}, Peter and {Harrington}, Joseph and {Hu}, Renyu and {Kendrew}, Sarah and {Knutson}, Heather A. and {Lagage}, Pierre-Olivier and {Leconte}, J{\'e}r{\'e}my and {Lendl}, Monika and {MacDonald}, Ryan J. and {May}, E.~M. and {Miguel}, Yamila and {Molaverdikhani}, Karan and {Moses}, Julianne I. and {Murray}, Catriona Anne and {Nehring}, Molly and {Nikolov}, Nikolay K. and {Petit dit de la Roche}, D.~J.~M. and {Radica}, Michael and {Roy}, Pierre-Alexis and {Stassun}, Keivan G. and {Taylor}, Jake and {Waalkes}, William C. and {Wachiraphan}, Patcharapol and {Welbanks}, Luis and {Wheatley}, Peter J. and {Aggarwal}, Keshav and {Alam}, Munazza K. and {Banerjee}, Agnibha and {Barstow}, Joanna K. and {Blecic}, Jasmina and {Casewell}, S.~L. and {Changeat}, Quentin and {Chubb}, K.~L. and {Col{\'o}n}, Knicole D. and {Coulombe}, Louis-Philippe and {Daylan}, Tansu and {de Val-Borro}, Miguel and {Decin}, Leen and {Dos Santos}, Leonardo A. and {Flagg}, Laura and {France}, Kevin and {Fu}, Guangwei and {Garc{\'\i}a Mu{\~n}oz}, A. and {Gizis}, John E. and {Glidden}, Ana and {Grant}, David and {Heng}, Kevin and {Henning}, Thomas and {Hong}, Yu-Cian and {Inglis}, Julie and {Iro}, Nicolas and {Kataria}, Tiffany and {Komacek}, Thaddeus D. and {Krick}, Jessica E. and {Lee}, Elspeth K.~H. and {Lewis}, Nikole K. and {Lillo-Box}, Jorge and {Lustig-Yaeger}, Jacob and {Mancini}, Luigi and {Mandell}, Avi M. and {Mansfield}, Megan and {Marley}, Mark S. and {Mikal-Evans}, Thomas and {Morello}, Giuseppe and {Nixon}, Matthew C. and {Ortiz Ceballos}, Kevin and {Piette}, Anjali A.~A. and {Powell}, Diana and {Rackham}, Benjamin V. and {Ramos-Rosado}, Lakeisha and {Rauscher}, Emily and {Redfield}, Seth and {Rogers}, Laura K. and {Roman}, Michael T. and {Roudier}, Gael M. and {Scarsdale}, Nicholas and {Shkolnik}, Evgenya L. and {Southworth}, John and {Spake}, Jessica J. and {Steinrueck}, Maria E. and {Tan}, Xianyu and {Teske}, Johanna K. and {Tremblin}, Pascal and {Tsai}, Shang-Min and {Tucker}, Gregory S. and {Turner}, Jake D. and {Valenti}, Jeff A. and {Venot}, Olivia and {Waldmann}, Ingo P. and {Wallack}, Nicole L. and {Zhang}, Xi and {Zieba}, Sebastian},
        title = "{Identification of carbon dioxide in an exoplanet atmosphere}",
      journal = {\nat},
         year = 2023,
        month = feb,
       volume = {614},
       number = {7949},
        pages = {649-652},
          doi = {10.1038/s41586-022-05269-w},
archivePrefix = {arXiv},
       eprint = {2208.11692},
 primaryClass = {astro-ph.EP},
       adsurl = {https://ui.adsabs.harvard.edu/abs/2023Natur.614..649J}
}

@ARTICLE{madhu24,
       author = {{Madhusudhan}, Nikku},
        title = "{The Hycean Paradigm in the Search for Life Elsewhere}",
      journal = {arXiv e-prints},
         year = 2024,
        month = jun,
          eid = {arXiv:2406.12794},
        pages = {arXiv:2406.12794},
          doi = {10.48550/arXiv.2406.12794},
archivePrefix = {arXiv},
       eprint = {2406.12794},
 primaryClass = {astro-ph.EP},
       adsurl = {https://ui.adsabs.harvard.edu/abs/2024arXiv240612794M}
}

@ARTICLE{seager10,
       author = {{Seager}, Sara and {Deming}, Drake},
        title = "{Exoplanet Atmospheres}",
      journal = {\araa},
         year = 2010,
        month = sep,
       volume = {48},
        pages = {631-672},
          doi = {10.1146/annurev-astro-081309-130837},
archivePrefix = {arXiv},
       eprint = {1005.4037},
 primaryClass = {astro-ph.EP},
       adsurl = {https://ui.adsabs.harvard.edu/abs/2010ARA&A..48..631S}
}

@ARTICLE{JF18,
       author = {{Fortney}, Jonathan J.},
        title = "{Modeling Exoplanetary Atmospheres: An Overview}",
      journal = {arXiv e-prints},
         year = 2018,
        month = apr,
          eid = {arXiv:1804.08149},
        pages = {arXiv:1804.08149},
          doi = {10.48550/arXiv.1804.08149},
archivePrefix = {arXiv},
       eprint = {1804.08149},
 primaryClass = {astro-ph.EP},
       adsurl = {https://ui.adsabs.harvard.edu/abs/2018arXiv180408149F}
}

@INCOLLECTION{JF21,
       author = {{Fortney}, Jonathan J. and {Barstow}, Joanna K. and {Madhusudhan}, Nikku},
        title = "{Atmospheric Modeling and Retrieval}",
    booktitle = {ExoFrontiers; Big Questions in Exoplanetary Science},
         year = 2021,
       editor = {{Madhusudhan}, Nikku},
        pages = {17-1},
          doi = {10.1088/2514-3433/abfa8fch17},
       adsurl = {https://ui.adsabs.harvard.edu/abs/2021exbi.book...17F}
}

@ARTICLE{Heng15,
       author = {{Heng}, Kevin and {Showman}, Adam P.},
        title = "{Atmospheric Dynamics of Hot Exoplanets}",
      journal = {Annual Review of Earth and Planetary Sciences},
         year = 2015,
        month = may,
       volume = {43},
        pages = {509-540},
          doi = {10.1146/annurev-earth-060614-105146},
archivePrefix = {arXiv},
       eprint = {1407.4150},
 primaryClass = {astro-ph.EP},
       adsurl = {https://ui.adsabs.harvard.edu/abs/2015AREPS..43..509H}
}

@ARTICLE{PICASO3.0,
       author = {{Mukherjee}, Sagnick and {Batalha}, Natasha E. and {Fortney}, Jonathan J. and {Marley}, Mark S.},
        title = "{PICASO 3.0: A One-dimensional Climate Model for Giant Planets and Brown Dwarfs}",
      journal = {\apj},
         year = 2023,
        month = jan,
       volume = {942},
       number = {2},
          eid = {71},
        pages = {71},
          doi = {10.3847/1538-4357/ac9f48},
archivePrefix = {arXiv},
       eprint = {2208.07836},
 primaryClass = {astro-ph.EP},
       adsurl = {https://ui.adsabs.harvard.edu/abs/2023ApJ...942...71M}
}

@ARTICLE{ATMO1,
       author = {{Tremblin}, P. and {Amundsen}, D.~S. and {Mourier}, P. and {Baraffe}, I. and {Chabrier}, G. and {Drummond}, B. and {Homeier}, D. and {Venot}, O.},
        title = "{Fingering Convection and Cloudless Models for Cool Brown Dwarf Atmospheres}",
      journal = {\apjl},
         year = 2015,
        month = may,
       volume = {804},
       number = {1},
          eid = {L17},
        pages = {L17},
          doi = {10.1088/2041-8205/804/1/L17},
archivePrefix = {arXiv},
       eprint = {1504.03334},
 primaryClass = {astro-ph.SR},
       adsurl = {https://ui.adsabs.harvard.edu/abs/2015ApJ...804L..17T}
}

@ARTICLE{ATMO2,
       author = {{Phillips}, M.~W. and {Tremblin}, P. and {Baraffe}, I. and {Chabrier}, G. and {Allard}, N.~F. and {Spiegelman}, F. and {Goyal}, J.~M. and {Drummond}, B. and {H{\'e}brard}, E.},
        title = "{A new set of atmosphere and evolution models for cool T-Y brown dwarfs and giant exoplanets}",
      journal = {\aap},
         year = 2020,
        month = may,
       volume = {637},
          eid = {A38},
        pages = {A38},
          doi = {10.1051/0004-6361/201937381},
archivePrefix = {arXiv},
       eprint = {2003.13717},
 primaryClass = {astro-ph.SR},
       adsurl = {https://ui.adsabs.harvard.edu/abs/2020A&A...637A..38P}
}

@article{VULCAN,
doi = {10.3847/1538-4357/ac29bc},
url = {https://dx.doi.org/10.3847/1538-4357/ac29bc},
year = {2021},
month = {dec},
publisher = {The American Astronomical Society},
volume = {923},
number = {2},
pages = {264},
author = {Shang-Min Tsai and Matej Malik and Daniel Kitzmann and James R. Lyons and Alexander Fateev and Elspeth Lee and Kevin Heng},
title = {A Comparative Study of Atmospheric Chemistry with VULCAN},
journal = {The Astrophysical Journal}
}

@article{THOR+HELIOS,
    author = {Deitrick, Russell and Heng, Kevin and Schroffenegger, Urs and Kitzmann, Daniel and Grimm, Simon L and Malik, Matej and Mendonça, João M and Morris, Brett M},
    title = "{The THOR + HELIOS general circulation model: multiwavelength radiative transfer with accurate scattering by clouds/hazes}",
    journal = {Monthly Notices of the Royal Astronomical Society},
    volume = {512},
    number = {3},
    pages = {3759-3787},
    year = {2022},
    month = {03},
    issn = {0035-8711},
    doi = {10.1093/mnras/stac680},
    url = {https://doi.org/10.1093/mnras/stac680},
    eprint = {https://academic.oup.com/mnras/article-pdf/512/3/3759/43286923/stac680.pdf},
}

@ARTICLE{Helios,
       author = {{Malik}, Matej and {Grosheintz}, Luc and {Mendon{\c{c}}a}, Jo{\~a}o M. and {Grimm}, Simon L. and {Lavie}, Baptiste and {Kitzmann}, Daniel and {Tsai}, Shang-Min and {Burrows}, Adam and {Kreidberg}, Laura and {Bedell}, Megan and {Bean}, Jacob L. and {Stevenson}, Kevin B. and {Heng}, Kevin},
        title = "{HELIOS: An Open-source, GPU-accelerated Radiative Transfer Code for Self-consistent Exoplanetary Atmospheres}",
      journal = {\aj},
         year = 2017,
        month = feb,
       volume = {153},
       number = {2},
          eid = {56},
        pages = {56},
          doi = {10.3847/1538-3881/153/2/56},
archivePrefix = {arXiv},
       eprint = {1606.05474},
 primaryClass = {astro-ph.EP},
       adsurl = {https://ui.adsabs.harvard.edu/abs/2017AJ....153...56M}
}

@ARTICLE{Fastchem,
       author = {{Stock}, Joachim W. and {Kitzmann}, Daniel and {Patzer}, A. Beate C.},
        title = "{FASTCHEM 2 : an improved computer program to determine the gas-phase chemical equilibrium composition for arbitrary element distributions}",
      journal = {\mnras},
         year = 2022,
        month = dec,
       volume = {517},
       number = {3},
        pages = {4070-4080},
          doi = {10.1093/mnras/stac2623},
archivePrefix = {arXiv},
       eprint = {2206.08247},
 primaryClass = {astro-ph.EP},
       adsurl = {https://ui.adsabs.harvard.edu/abs/2022MNRAS.517.4070S}
}

@ARTICLE{Solar_2009,
       author = {{Asplund}, Martin and {Grevesse}, Nicolas and {Sauval}, A. Jacques and {Scott}, Pat},
        title = "{The Chemical Composition of the Sun}",
      journal = {\araa},
         year = 2009,
        month = sep,
       volume = {47},
       number = {1},
        pages = {481-522},
          doi = {10.1146/annurev.astro.46.060407.145222},
archivePrefix = {arXiv},
       eprint = {0909.0948},
 primaryClass = {astro-ph.SR},
       adsurl = {https://ui.adsabs.harvard.edu/abs/2009ARA&A..47..481A}
}

@ARTICLE{petit,
       author = {{Molli{\`e}re}, P. and {Wardenier}, J.~P. and {van Boekel}, R. and {Henning}, Th. and {Molaverdikhani}, K. and {Snellen}, I.~A.~G.},
        title = "{petitRADTRANS. A Python radiative transfer package for exoplanet characterization and retrieval}",
      journal = {\aap},
         year = 2019,
        month = jul,
       volume = {627},
          eid = {A67},
        pages = {A67},
          doi = {10.1051/0004-6361/201935470},
archivePrefix = {arXiv},
       eprint = {1904.11504},
 primaryClass = {astro-ph.EP},
       adsurl = {https://ui.adsabs.harvard.edu/abs/2019A&A...627A..67M}
}

@ARTICLE{gCMCRT,
       author = {{Lee}, Elspeth K.~H. and {Wardenier}, Joost P. and {Prinoth}, Bibiana and {Parmentier}, Vivien and {Grimm}, Simon L. and {Baeyens}, Robin and {Carone}, Ludmila and {Christie}, Duncan and {Deitrick}, Russell and {Kitzmann}, Daniel and {Mayne}, Nathan and {Roman}, Michael and {Thorsbro}, Brian},
        title = "{3D Radiative Transfer for Exoplanet Atmospheres. gCMCRT: A GPU-accelerated MCRT Code}",
      journal = {\apj},
         year = 2022,
        month = apr,
       volume = {929},
       number = {2},
          eid = {180},
        pages = {180},
          doi = {10.3847/1538-4357/ac61d6},
archivePrefix = {arXiv},
       eprint = {2110.15640},
 primaryClass = {astro-ph.EP},
       adsurl = {https://ui.adsabs.harvard.edu/abs/2022ApJ...929..180L}
}

@article{HDhotspot1,
    author = {Pont, F. and Sing, D. K. and Gibson, N. P. and Aigrain, S. and Henry, G. and Husnoo, N.},
    title = "{The prevalence of dust on the exoplanet HD 189733b from Hubble and Spitzer observations}",
    journal = {Monthly Notices of the Royal Astronomical Society},
    volume = {432},
    number = {4},
    pages = {2917-2944},
    year = {2013},
    month = {05},
    issn = {0035-8711},
    doi = {10.1093/mnras/stt651},
    url = {https://doi.org/10.1093/mnras/stt651},
    eprint = {https://academic.oup.com/mnras/article-pdf/432/4/2917/18613540/stt651.pdf},
}

@article{HDhotspot2,
	author = {{de Wit, J.} and {Gillon, M.} and {Demory, B.-O.} and {Seager, S.}},
	title = {Towards consistent mapping of distant worlds:  secondary-eclipse scanning of the exoplanet HD733b},
	DOI= "10.1051/0004-6361/201219060",
	url= "https://doi.org/10.1051/0004-6361/201219060",
	journal = {A \& A},
	year = 2012,
	volume = 548,
	pages = "A128",
	month = "",
}

@ARTICLE{picaso,
       author = {{Batalha}, Natasha E. and {Marley}, Mark S. and {Lewis}, Nikole K. and {Fortney}, Jonathan J.},
        title = "{Exoplanet Reflected-light Spectroscopy with PICASO}",
      journal = {\apj},
         year = 2019,
        month = jun,
       volume = {878},
       number = {1},
          eid = {70},
        pages = {70},
          doi = {10.3847/1538-4357/ab1b51},
archivePrefix = {arXiv},
       eprint = {1904.09355},
 primaryClass = {astro-ph.EP},
       adsurl = {https://ui.adsabs.harvard.edu/abs/2019ApJ...878...70B}
}

@ARTICLE{Phoenix_rt,
       author = {{Barman}, Travis S. and {Hauschildt}, Peter H. and {Allard}, France},
        title = "{Irradiated Planets}",
      journal = {\apj},
         year = 2001,
        month = aug,
       volume = {556},
       number = {2},
        pages = {885-895},
          doi = {10.1086/321610},
archivePrefix = {arXiv},
       eprint = {astro-ph/0104262},
 primaryClass = {astro-ph},
       adsurl = {https://ui.adsabs.harvard.edu/abs/2001ApJ...556..885B}
}

@article{Showman_2009,
	doi = {10.1088/0004-637x/699/1/564},
  
	url = {https://doi.org/10.1088%2F0004-637x%2F699%2F1%2F564},
  
	year = 2009,
	month = {jun},
  
	publisher = {American Astronomical Society},
  
	volume = {699},
  
	number = {1},
  
	pages = {564--584},
  
	author = {Adam P. Showman and Jonathan J. Fortney and Yuan Lian and Mark S. Marley and Richard S. Freedman and Heather A. Knutson and David Charbonneau},
  
	title = {{ATMOSPHERIC} {CIRCULATION} {OF} {HOT} {JUPITERS}: {COUPLED} {RADIATIVE}-{DYNAMICAL} {GENERAL} {CIRCULATION} {MODEL} {SIMULATIONS} {OF} {HD} 189733b and {HD} 209458b},
  
	journal = {The Astrophysical Journal}
}

@ARTICLE{exofms,
       author = {{Lee}, Elspeth K.~H. and {Parmentier}, Vivien and {Hammond}, Mark and {Grimm}, Simon L. and {Kitzmann}, Daniel and {Tan}, Xianyu and {Tsai}, Shang-Min and {Pierrehumbert}, Raymond T.},
        title = "{Simulating gas giant exoplanet atmospheres with EXO-FMS: comparing semigrey, picket fence, and correlated-k radiative-transfer schemes}",
      journal = {\mnras},
         year = 2021,
        month = sep,
       volume = {506},
       number = {2},
        pages = {2695-2711},
          doi = {10.1093/mnras/stab1851},
archivePrefix = {arXiv},
       eprint = {2106.11664},
 primaryClass = {astro-ph.EP},
       adsurl = {https://ui.adsabs.harvard.edu/abs/2021MNRAS.506.2695L}
}

@ARTICLE{UM,
       author = {{Mayne}, Nathan J. and {Baraffe}, Isabelle and {Acreman}, David M. and {Smith}, Chris and {Browning}, Matthew K. and {Sk{\r{a}}lid Amundsen}, David and {Wood}, Nigel and {Thuburn}, John and {Jackson}, David R.},
        title = "{The unified model, a fully-compressible, non-hydrostatic, deep atmosphere global circulation model, applied to hot Jupiters. ENDGame for a HD 209458b test case}",
      journal = {\aap},
         year = 2014,
        month = jan,
       volume = {561},
          eid = {A1},
        pages = {A1},
          doi = {10.1051/0004-6361/201322174},
       adsurl = {https://ui.adsabs.harvard.edu/abs/2014A&A...561A...1M}
}

@INPROCEEDINGS{trap3d,
       author = {{Quirino}, Diogo and {Gilli}, Gabriella and {Navarro}, Thomas and {Turbet}, Martin and {Fauchez}, Thomas and {Machado}, Pedro},
        title = "{3D Climate modelling of TRAPPIST-1 c with a Venus-like atmosphere: large-scale circulation and observational prospects}",
    booktitle = {EGU General Assembly Conference Abstracts},
         year = 2022,
       series = {EGU General Assembly Conference Abstracts},
        month = may,
          eid = {EGU22-8185},
        pages = {EGU22-8185},
          doi = {10.5194/egusphere-egu22-8185},
       adsurl = {https://ui.adsabs.harvard.edu/abs/2022EGUGA..24.8185Q}
}

@INPROCEEDINGS{k2183d,
       author = {{Innes}, Hamish I. and {Pierrehumbert}, Raymond T.},
        title = "{Characterising the circulation of temperate sub-Neptunes using 3D GCM simulations}",
    booktitle = {Bulletin of the American Astronomical Society},
         year = 2022,
       volume = {54},
        month = jun,
          eid = {102.210},
        pages = {102.210},
       adsurl = {https://ui.adsabs.harvard.edu/abs/2022BAAS...54e.210I}
}

@ARTICLE{806063d,
       author = {{Tsai}, Shang-Min and {Steinrueck}, Maria and {Parmentier}, Vivien and {Lewis}, Nikole and {Pierrehumbert}, Raymond},
        title = "{The climate and compositional variation of the highly eccentric planet HD 80606 b - the rise and fall of carbon monoxide and elemental sulfur}",
      journal = {\mnras},
         year = 2023,
        month = apr,
       volume = {520},
       number = {3},
        pages = {3867-3886},
          doi = {10.1093/mnras/stad214},
       adsurl = {https://ui.adsabs.harvard.edu/abs/2023MNRAS.520.3867T}
}

@ARTICLE{hd189733b3d,
       author = {{Steinrueck}, Maria E. and {Koskinen}, Tommi and {Lavvas}, Panayotis and {Parmentier}, Vivien and {Zieba}, Sebastian and {Tan}, Xianyu and {Zhang}, Xi and {Kreidberg}, Laura},
        title = "{Photochemical hazes dramatically alter temperature structure and atmospheric circulation in 3D simulations of hot Jupiters}",
      journal = {arXiv e-prints},
         year = 2023,
        month = may,
          eid = {arXiv:2305.09654},
        pages = {arXiv:2305.09654},
          doi = {10.48550/arXiv.2305.09654},
archivePrefix = {arXiv},
       eprint = {2305.09654},
 primaryClass = {astro-ph.EP},
       adsurl = {https://ui.adsabs.harvard.edu/abs/2023arXiv230509654S}
}

@ARTICLE{Wasp39b3d,
       author = {{Tsai}, Shang-Min and {Moses}, Julianne I. and {Powell}, Diana and {Lee}, Elspeth K.~H.},
        title = "{Day-night transport induced chemistry and clouds on WASP-39b I: Gas-phase composition}",
      journal = {arXiv e-prints},
         year = 2023,
        month = may,
          eid = {arXiv:2305.19403},
        pages = {arXiv:2305.19403},
          doi = {10.48550/arXiv.2305.19403},
archivePrefix = {arXiv},
       eprint = {2305.19403},
 primaryClass = {astro-ph.EP},
       adsurl = {https://ui.adsabs.harvard.edu/abs/2023arXiv230519403T}
}

@ARTICLE{HD189water1,
       author = {{Grillmair}, Carl J. and {Burrows}, Adam and {Charbonneau}, David and {Armus}, Lee and {Stauffer}, John and {Meadows}, Victoria and {van Cleve}, Jeffrey and {von Braun}, Kaspar and {Levine}, Deborah},
        title = "{Strong water absorption in the dayside emission spectrum of the planet HD189733b}",
      journal = {\nat},
         year = 2008,
        month = dec,
       volume = {456},
       number = {7223},
        pages = {767-769},
          doi = {10.1038/nature07574},
archivePrefix = {arXiv},
       eprint = {0901.4774},
 primaryClass = {astro-ph.EP},
       adsurl = {https://ui.adsabs.harvard.edu/abs/2008Natur.456..767G}
}

@article{HD189water2,
	doi = {10.3847/1538-3881/ac1f8e},
  
	url = {https://doi.org/10.3847%2F1538-3881%2Fac1f8e},
  
	year = 2021,
	month = {nov},
  
	publisher = {American Astronomical Society},
  
	volume = {162},
  
	number = {6},
  
	pages = {233},
  
	author = {Anne Boucher and Antoine Darveau-Bernier and Stefan Pelletier and David Lafreni{\`{e}
}re and {\'{E}}tienne Artigau and Neil J. Cook and Romain Allart and Michael Radica and Ren{\'{e}} Doyon and Björn Benneke and Luc Arnold and Xavier Bonfils and Vincent Bourrier and Ryan Cloutier and Jo{\~{a}}o Gomes da Silva and Emily Deibert and Xavier Delfosse and Jean-Fran{\c{c}}ois Donati and David Ehrenreich and Pedro Figueira and Thierry Forveille and Pascal Fouqu{\'{e}} and Jonathan Gagn{\'{e}} and Eric Gaidos and Guillaume H{\'{e}}brard and Ray Jayawardhana and Baptiste Klein and Christophe Lovis and Jorge H. C. Martins and Eder Martioli and Claire Moutou and Nuno C. Santos},
  
	title = {Characterizing Exoplanetary Atmospheres at High Resolution with {SPIRou}: Detection of Water on {HD} 189733 b},
  
	journal = {The Astronomical Journal}
}

@ARTICLE{HD189hcyn,
       author = {{Cabot}, Samuel H.~C. and {Madhusudhan}, Nikku and {Hawker}, George A. and {Gandhi}, Siddharth},
        title = "{On the robustness of analysis techniques for molecular detections using high-resolution exoplanet spectroscopy}",
      journal = {\mnras},
         year = 2019,
        month = feb,
       volume = {482},
       number = {4},
        pages = {4422-4436},
          doi = {10.1093/mnras/sty2994},
archivePrefix = {arXiv},
       eprint = {1811.05978},
 primaryClass = {astro-ph.EP},
       adsurl = {https://ui.adsabs.harvard.edu/abs/2019MNRAS.482.4422C}
}

@ARTICLE{HD189na,
       author = {{Wyttenbach}, A. and {Ehrenreich}, D. and {Lovis}, C. and {Udry}, S. and {Pepe}, F.},
        title = "{Spectrally resolved detection of sodium in the atmosphere of HD 189733b with the HARPS spectrograph}",
      journal = {\aap},
         year = 2015,
        month = may,
       volume = {577},
          eid = {A62},
        pages = {A62},
          doi = {10.1051/0004-6361/201525729},
archivePrefix = {arXiv},
       eprint = {1503.05581},
 primaryClass = {astro-ph.EP},
       adsurl = {https://ui.adsabs.harvard.edu/abs/2015A&A...577A..62W}
}

@article{HD189k,
    author = {Keles, Engin and Mallonn, Matthias and von Essen, Carolina and Carroll, Thorsten A and Alexoudi, Xanthippi and Pino, Lorenzo and Ilyin, Ilya and Poppenhäger, Katja and Kitzmann, Daniel and Nascimbeni, Valerio and Turner, Jake D and Strassmeier, Klaus G},
    title = "{The potassium absorption on HD189733b and HD209458b}",
    journal = {Monthly Notices of the Royal Astronomical Society: Letters},
    volume = {489},
    number = {1},
    pages = {L37-L41},
    year = {2019},
    month = {08},
    issn = {1745-3925},
    doi = {10.1093/mnrasl/slz123},
    url = {https://doi.org/10.1093/mnrasl/slz123},
    eprint = {https://academic.oup.com/mnrasl/article-pdf/489/1/L37/29208449/slz123.pdf},
}

@article{ HD189co,
	author = {{de Kok, R.J.} and {Brogi, M.} and {Snellen, I.A.G.} and {Birkby, J.} and {Albrecht, S.} and {de Mooij, E.J.W.}},
	title = {Detection of carbon monoxide in the high-resolution day-side spectrum of the exoplanet HD 189733b},
	DOI= "10.1051/0004-6361/201321381",
	url= "https://doi.org/10.1051/0004-6361/201321381",
	journal = {A\&A},
	year = 2013,
	volume = 554,
	pages = "A82",
	month = "",
}

@article{3dhdspitzer,
	doi = {10.1051/0004-6361/201219060},
  
	url = {https://doi.org/10.1051%2F0004-6361%2F201219060},
  
	year = 2012,
	month = {dec},
  
	publisher = {{EDP} Sciences},
  
	volume = {548},
  
	pages = {A128},
  
	author = {J. de Wit and M. Gillon and B.-O. Demory and S. Seager},
  
	title = {Towards consistent mapping of distant worlds: secondary-eclipse scanning of the exoplanet {HD}{\hspace{0.167em}
}189733b},
  
	journal = {Astronomy \& Astrophysics}
}

@article{Pandexo,
	doi = {10.1088/1538-3873/aa65b0},
  
	url = {https://doi.org/10.1088%2F1538-3873%2Faa65b0},
  
	year = 2017,
	month = {apr},
  
	publisher = {{IOP} Publishing},
  
	volume = {129},
  
	number = {976},
  
	pages = {064501},
  
	author = {Natasha E. Batalha and Avi Mandell and Klaus Pontoppidan and Kevin B. Stevenson and Nikole K. Lewis and Jason Kalirai and Nick Earl and Thomas Greene and Loïc Albert and Louise D. Nielsen},
  
	title = {{PandExo}: A Community Tool for Transiting Exoplanet Science with$\less$i$\greater${JWST}$\less$/i$\greater${\&}amp$\mathsemicolon$$\less$i$\greater${HST}$\less$/i$\greater$},
  
	journal = {Publications of the Astronomical Society of the Pacific}
}

@ARTICLE{SNR,
       author = {{Lustig-Yaeger}, Jacob and {Meadows}, Victoria S. and {Lincowski}, Andrew P.},
        title = "{The Detectability and Characterization of the TRAPPIST-1 Exoplanet Atmospheres with JWST}",
      journal = {\aj},
         year = 2019,
        month = jul,
       volume = {158},
       number = {1},
          eid = {27},
        pages = {27},
          doi = {10.3847/1538-3881/ab21e0},
archivePrefix = {arXiv},
       eprint = {1905.07070},
 primaryClass = {astro-ph.EP},
       adsurl = {https://ui.adsabs.harvard.edu/abs/2019AJ....158...27L}
}

@article{SNR2,
	doi = {10.3847/psj/ac6cf1},
  
	url = {https://doi.org/10.3847%2Fpsj%2Fac6cf1},
  
	year = 2022,
	month = {sep},
  
	publisher = {American Astronomical Society},
  
	volume = {3},
  
	number = {9},
  
	pages = {213},
  
	author = {Thomas J. Fauchez and Geronimo L. Villanueva and Denis E. Sergeev and Martin Turbet and Ian A. Boutle and Kostas Tsigaridis and Michael J. Way and Eric T. Wolf and Shawn D. Domagal-Goldman and Fran{\c{c}
}ois Forget and Jacob Haqq-Misra and Ravi K. Kopparapu and James Manners and Nathan J. Mayne},
  
	title = {The {TRAPPIST}-1 Habitable Atmosphere Intercomparison ({THAI}). {III}. Simulated Observables{\textemdash}the Return of the Spectrum},
  
	journal = {The Planetary Science Journal}
}

@MISC{2019ascl.soft03014Z,
author = {{Zhang}, Michael and {Chachan}, Yayaati},
title = {PLATON: PLanetary Atmospheric Transmission for Observer Noobs},
howpublished = {Astrophysics Source Code Library, record ascl:1903.014},
year = 2019,
month = mar,
eid = {ascl:1903.014},
pages = {ascl:1903.014},
archivePrefix = {ascl},
eprint = {1903.014},
adsurl = {https://ui.adsabs.harvard.edu/abs/2019ascl.soft03014Z}
}

@article{Tremblin2015,
adsurl = {http://adsabs.harvard.edu/abs/2015ApJ...804L..17T},
archiveprefix = {arXiv},
author = {{Tremblin}, P. and {Amundsen}, D.~S. and {Mourier}, P. and
{Baraffe}, I. and {Chabrier}, G. and {Drummond}, B. and {Homeier}, D. and {Venot},
O.},
doi = {10.1088/2041-8205/804/1/L17},
eid = {L17},
eprint = {1504.03334},
journal = {\apjl},
month = may,
pages = {L17},
primaryclass = {astro-ph.SR},
title = {{Fingering Convection and Cloudless Models for Cool Brown Dwarf
Atmospheres}},
volume = 804,
year = 2015}

@article{Drummond2016,
adsurl = {http://adsabs.harvard.edu/abs/2016A%26A...594A..69D},
archiveprefix = {arXiv},
author = {{Drummond}, B. and {Tremblin}, P. and {Baraffe}, I. and
{Amundsen}, D.~S. and {Mayne}, N.~J. and {Venot}, O. and {Goyal}, J.},
doi = {10.1051/0004-6361/201628799},
eid = {A69},
eprint = {1607.04062},
journal = {\aap},
month = oct,
pages = {A69},
primaryclass = {astro-ph.EP},
title = {{The effects of consistent chemical kinetics calculations on the
pressure-temperature profiles and emission spectra of hot Jupiters}},
volume = 594,
year = 2016}

@article{Goyal2018,
adsurl = {https://ui.adsabs.harvard.edu/#abs/2018MNRAS.474.5158G},
author = {{Goyal}, Jayesh M. and {Mayne}, Nathan and {Sing}, David K. and
{Drummond}, Benjamin and {Tremblin}, Pascal and {Amundsen}, David S. and
{Evans}, Thomas and {Carter}, Aarynn L. and {Spake}, Jessica and {Baraffe},
Isabelle and {Nikolov}, Nikolay and {Manners}, James and {Chabrier}, Gilles and
{Hebrard}, Eric},
doi = {10.1093/mnras/stx3015},
journal = {\mnras},
month = Mar,
pages = {5158-5185},
title = {{A library of ATMO forward model transmission spectra for hot Jupiter
exoplanets}},
volume = {474},
year = 2018}

@article{Boyajian_2014,
   title={Stellar diameters and temperatures – VI. High angular resolution measurements of the transiting exoplanet host stars HD 189733 and HD 209458 and implications for models of cool dwarfs},
   volume={447},
   ISSN={0035-8711},
   url={http://dx.doi.org/10.1093/mnras/stu2502},
   DOI={10.1093/mnras/stu2502},
   number={1},
   journal={Monthly Notices of the Royal Astronomical Society},
   publisher={Oxford University Press (OUP)},
   author={Boyajian, Tabetha and von Braun, Kaspar and Feiden, Gregory A. and Huber, Daniel and Basu, Sarbani and Demarque, Pierre and Fischer, Debra A. and Schaefer, Gail and Mann, Andrew W. and White, Timothy R. and Maestro, Vicente and Brewer, John and Lamell, C. Brooke and Spada, Federico and López-Morales, Mercedes and Ireland, Michael and Farrington, Chris and van Belle, Gerard T. and Kane, Stephen R. and Jones, Jeremy and ten Brummelaar, Theo A. and Ciardi, David R. and McAlister, Harold A. and Ridgway, Stephen and Goldfinger, P. J. and Turner, Nils H. and Sturmann, Laszlo},
   year={2014},
   month=dec, pages={846–857} }

@article{Deitrick_2020,
   title={THOR 2.0: Major Improvements to the Open-source General Circulation Model},
   volume={248},
   ISSN={1538-4365},
   url={http://dx.doi.org/10.3847/1538-4365/ab930e},
   DOI={10.3847/1538-4365/ab930e},
   number={2},
   journal={The Astrophysical Journal Supplement Series},
   publisher={American Astronomical Society},
   author={Deitrick, Russell and Mendonça, João M. and Schroffenegger, Urs and Grimm, Simon L. and Tsai, Shang-Min and Heng, Kevin},
   year={2020},
   month=jun, pages={30} }

@misc{fu2024hydrogensulfidemetalenrichedatmosphere,
      title={Hydrogen sulfide and metal-enriched atmosphere for a Jupiter-mass exoplanet}, 
      author={Guangwei Fu and Luis Welbanks and Drake Deming and Julie Inglis and Michael Zhang and Joshua Lothringer and Jegug Ih and Julianne I. Moses and Everett Schlawin and Heather A. Knutson and Gregory Henry and Thomas Greene and David K. Sing and Arjun B. Savel and Eliza M. -R. Kempton and Dana R. Louie and Michael Line and Matt Nixon},
      year={2024},
      eprint={2407.06163},
      archivePrefix={arXiv},
      primaryClass={astro-ph.EP},
      url={https://arxiv.org/abs/2407.06163}, 
}

@ARTICLE{2024ApJ...973L..41I,
       author = {{Inglis}, Julie and {Batalha}, Natasha E. and {Lewis}, Nikole K. and {Kataria}, Tiffany and {Knutson}, Heather A. and {Kilpatrick}, Brian M. and {Gagnebin}, Anna and {Mukherjee}, Sagnick and {Pettyjohn}, Maria M. and {Crossfield}, Ian J.~M. and {Foote}, Trevor O. and {Grant}, David and {Henry}, Gregory W. and {Lally}, Maura and {McKemmish}, Laura K. and {Sing}, David K. and {Wakeford}, Hannah R. and {Zapata Trujillo}, Juan C. and {Zellem}, Robert T.},
        title = "{Quartz Clouds in the Dayside Atmosphere of the Quintessential Hot Jupiter HD 189733 b}",
      journal = {\apjl},
         year = 2024,
        month = oct,
       volume = {973},
       number = {2},
          eid = {L41},
        pages = {L41},
          doi = {10.3847/2041-8213/ad725e},
archivePrefix = {arXiv},
       eprint = {2409.11395},
 primaryClass = {astro-ph.EP},
       adsurl = {https://ui.adsabs.harvard.edu/abs/2024ApJ...973L..41I}
}

@ARTICLE{2023A&A...672A.110L,
       author = {{Lee}, Elspeth K.~H. and {Tsai}, Shang-Min and {Hammond}, Mark and {Tan}, Xianyu},
        title = "{A mini-chemical scheme with net reactions for 3D general circulation models. II. 3D thermochemical modelling of WASP-39b and HD 189733b}",
      journal = {\aap},
         year = 2023,
        month = apr,
       volume = {672},
          eid = {A110},
        pages = {A110},
          doi = {10.1051/0004-6361/202245473},
archivePrefix = {arXiv},
       eprint = {2302.09525},
 primaryClass = {astro-ph.EP},
       adsurl = {https://ui.adsabs.harvard.edu/abs/2023A&A...672A.110L}
}

@ARTICLE{drummond2020,
       author = {{Drummond}, Benjamin and {H{\'e}brard}, Eric and {Mayne}, Nathan J. and {Venot}, Olivia and {Ridgway}, Robert J. and {Changeat}, Quentin and {Tsai}, Shang-Min and {Manners}, James and {Tremblin}, Pascal and {Abraham}, Nathan Luke and {Sing}, David and {Kohary}, Krisztian},
        title = "{Implications of three-dimensional chemical transport in hot Jupiter atmospheres: Results from a consistently coupled chemistry-radiation-hydrodynamics model}",
      journal = {\aap},
         year = 2020,
        month = apr,
       volume = {636},
          eid = {A68},
        pages = {A68},
          doi = {10.1051/0004-6361/201937153},
archivePrefix = {arXiv},
       eprint = {2001.11444},
 primaryClass = {astro-ph.EP},
       adsurl = {https://ui.adsabs.harvard.edu/abs/2020A&A...636A..68D}
}

@ARTICLE{zam2023,
       author = {{Zamyatina}, Maria and {H{\'e}brard}, Eric and {Drummond}, Benjamin and {Mayne}, Nathan J. and {Manners}, James and {Christie}, Duncan A. and {Tremblin}, Pascal and {Sing}, David K. and {Kohary}, Krisztian},
        title = "{Observability of signatures of transport-induced chemistry in clear atmospheres of hot gas giant exoplanets}",
      journal = {\mnras},
         year = 2023,
        month = feb,
       volume = {519},
       number = {2},
        pages = {3129-3153},
          doi = {10.1093/mnras/stac3432},
archivePrefix = {arXiv},
       eprint = {2211.09071},
 primaryClass = {astro-ph.EP},
       adsurl = {https://ui.adsabs.harvard.edu/abs/2023MNRAS.519.3129Z}
}

\end{document}